\documentclass[12pt]{iopart}
\font\cal=cmsy10 scaled 1200
       \newcommand{\Cc}{\mbox{\cal\symbol{'103}}}
       \newcommand{\Dc}{\mbox{\cal\symbol{'104}}}
       \newcommand{\Ec}{\mbox{\cal\symbol{'105}}}
       \newcommand{\Gc}{\mbox{\cal\symbol{'107}}}
       \newcommand{\Hc}{\mbox{\cal\symbol{'110}}}
       \newcommand{\Ic}{\mbox{\cal\symbol{'111}}}
       
       \newcommand{\Pc}{\mbox{\cal\symbol{'120}}}
       \newcommand{\Rc}{\mbox{\cal\symbol{'122}}}
\font\semibold=msbm10 scaled 1200

       \newcommand{\R}{\mbox{\semibold\symbol{'122}}}
\font\bfgreek=cmmib10 scaled 1200
       \newcommand{\bftheta}{\mbox{\bfgreek\symbol{'022}}}

\newcommand{\w}{\wedge}

\begin{document}

\title{Contact Equivalence Problem for Linear Parabolic Equations}

\author{Oleg I. Morozov}

\address{Department of Mathematics, Snezhinsk Physical and Technical Academy,

\noindent
Snezhinsk, 456776, Russia

\noindent
morozov{\symbol{64}}sfti.snz.ru}

\begin{abstract}
The moving coframe method is applied to solve the local equivalence problem
for the class of linear parabolic equations in two independent variables
under an action of the pseudo-group of contact transformations. The structure
equations and the complete sets of differential invariants for symmetry groups
are found. The solution of the equivalence problem is given in terms of these
invariants.
\end{abstract}


\section*{Introduction}
In this article we consider a local equivalence problem for the class
of linear second order parabolic equations
\begin{equation}
u_{xx} = T(t,x)\,u_t+X(t,x)\,u_x+U(t,x)\,u
\label{para}
\end{equation}
\noindent under a contact transformation pseudo-group. Two
equations are said to be equi\-va\-lent if there exists a contact
transformation mapping one equation to the other. \'Elie Cartan
developed a general method for solving equivalence problems,
\cite{Cartan1} - \cite{Cartan5}. The method provides an effective
means of computing complete systems of differential invariants and
associated invariant differential operators. The necessary and
sufficient condition for equivalence of two sub\-ma\-ni\-folds
under an action of a Lie pseudo-group is formulated in terms of
the dif\-fe\-ren\-ti\-al invariants. The invariants parameterize
the classifying manifold associated with given submanifolds.
Cartan's solution to the equivalence problem states that two
sub\-ma\-ni\-folds are (locally) equivalent if and only if their
classifying manifolds (locally) over\-lap. The sym\-met\-ry
clas\-si\-fi\-ca\-tion problem for classes of differential
equations is closely related to the problem of local equivalence:
symmetry groups and their Lie al\-ge\-bras of two equations are
necessarily isomorphic if these equations are equivalent, while
the converse statement is not true in general. The symmetry
analysis of linear second order parabolic equations (\ref{para})
is done by Sophus Lie, \cite[Vol.~3, pp~492-523]{Lie}. In
\cite[\S~9]{Ovsiannikov}, Ovsiannikov gives the finite defining
equation for the equivalence pseudo-group and the symmetry
classification in terms of a nor\-mal form $u_{xx}=u_t+H(t,x)\,u$
for equations (\ref{para}). In \cite{Ibragimov2002}, the Laplace
type semi-invariant, i.e., the function remaining unchanged under
a transformation $\overline{u}=\sigma(t,x)\,u$ for every
$\sigma(t,x)$, is found for the class (\ref{para}). This function
\begin{equation}\fl
K = (2\,T\,X\,X_x-X^2\,T_x+2\,T_x\,X_x+2\,T^2\,X_t-2\,T\,X_xx+4\,T\,U_x
-4\,U\,T_x)/(2\,T^4)
\label{Ibragimov_Kappa}
\end{equation}
\noindent
is not invariant under the full symmetry group of equation (\ref{para}).
In \cite{JM}, it is shown that equation (\ref{para}) is reducible to the heat
equation $u_{xx}=u_t$ under some contact trans\-for\-ma\-ti\-on if and only if
$\lambda = 0$, where
\[\fl
\lambda = (
8\,T^8\,K_{xx}+20\,T^7\,T_x\,K_x+12\,T^7\,T_{xx}\,K+288\,T^2\,T_x\,T_{xx}^2
+220\,T^2\,T_x\,T_{xxx}
\]
\[\fl\hspace{15pt}
-64\,T^3\,T_{xx}\,T_{xxx}-40\,T^3\,T_x\,T_{xxxx}
+4\,T^4\,T_{xxxxx}+4\,T^6\,T_{ttx}-8\,T^5\,T_{txx}+405\,T_x^5
\]
\[\fl\hspace{15pt}
-810\,T\,T_x^3\,T_{xx}+4\,T^4\,T_x\,T_t^2+4\,T^5\,T_x\,T_{tt}^2
+80\,T^2\,T_t\,T_x^3-4\,T^5\,T_t\,T_{tx}-80\,T^3\,T_x^2\,T_{tx}
\]
\begin{equation}
\fl\hspace{15pt}
+28\,T^4\,T_{tx}\,T_{xx}+36\,T^4\,T_x\,T_{txx}+8\,T^4\,T_t\,T_{xxx}
-64\,T^3\,T_t\,T_x\,T_{xx})/T^{10},
\label{JM_lambda}
\end{equation}
\noindent
and $K$ is defined by (\ref{Ibragimov_Kappa}).

In the present paper, we apply Cartan's equivalence method, \cite{Cartan1} -
\cite{Cartan5}, \cite{Gardner}, \cite{Olver95}, in its form developed by
Fels and Olver, \cite{FO,FO2}, to find all differential invariants of
sym\-met\-ry groups for equations (\ref{para}) and to solve the local contact
equivalence problem for equations from the class (\ref{para}) in terms of their
coefficients. Examples of computing structure for symmetry pseudo-groups of
partial differential equations via the method of \cite{FO,FO2} are given in
\cite{Morozov}. Unlike Lie's infinitesimal method, Cartan's approach allows
us to find differential invariants and invariant differential operators without
analysing over-determined systems of PDEs at all, and requires differentiation
and linear algebra operations only.

The paper is organized as follows. In Section 1, we begin with
some notation, and use Cartan's equivalence method to find the
invariant 1-forms and the structure equations for the pseudo-group
of contact transformations on the bundle of second-order jets. In
Section 2, we briefly describe the approach to computing symmetry
groups of differential equations via the moving coframe method of
Fels and Olver. In Section 3, the method is applied to the class
of parabolic equations (\ref{para}). Finally, we make some
concluding remarks.

\section{Pseudo-group of contact transformations}

In this paper, all considerations are of local nature, and all mappings are
real analytic. Suppose $\Ec = \R^n \times \R \rightarrow \R^n$ is a trivial
bundle with the local base coordinates $(x^1,...,x^n)$ and the local fibre
coordinate $u$; then by $J^2(\Ec)$ denote the bundle of the second-order jets
of sections of $\Ec$, with the local coordinates $(x^i,u,p_i,p_{ij})$,
$i,j\in\{1,...,n\}$, $i \le j$. For every local section $(x^i,f(x))$ of $\Ec$,
the corresponding 2-jet $(x^i,f(x),\partial f(x)/\partial x^i,
\partial^2 f(x)/\partial x^i\partial x^j)$ is denoted by $j_2(f)$.
A differential 1-form $\vartheta$ on $J^2(\Ec)$ is called a {\it contact
form}, if it is annihilated by all 2-jets of local sections:
$j_2(f)^{*}\vartheta = 0$. In the local coordinates every contact 1-form is a
linear combination of the forms  $\vartheta_0 = du - p_{i}\,dx^i$,
$\vartheta_i = dp_i - p_{ij}\,dx^j$, $i, j \in \{1,...,n\}$, $p_{ji} = p_{ij}$
(here and later we use the Einstein summation convention, so
$p_i\,dx^i = \sum_{i=1}^{n}\,p_i\,dx^i$, etc.)
A local diffeomorphism
\begin{equation}
\Delta : J^2(\Ec) \rightarrow J^2(\Ec),
\qquad
\Delta : (x^i,u,p_i,p_{ij}) \mapsto
(\overline{x}^i,\overline{u},\overline{p}_i,\overline{p}_{ij}),
\label{Delta}
\end{equation}
\noindent
is called a {\it contact transformation}, if for every contact 1-form
$\vartheta$, the form $\Delta^{*}\overline{\vartheta}$ is also contact.
To obtain a collection of invariant 1-forms for the pseudo-group of
contact transformations on $J^2(\Ec)$, we apply Cartan's method of equivalence,
\cite{Cartan5,Olver95}. For this, take the coframe
$\{\vartheta_0, \vartheta_i, dx^i, dp_{ij}\,\vert\,i,j\in\{1,...,n\}, i\le j\}$
on $J^2(\Ec)$. A contact transformation (\ref{Delta}) acts on this coframe in
the following manner:
\[
\Delta^{*}\,
\left(
\begin{array}{c}
\overline{\vartheta}_0\\
\overline{\vartheta}_i\\
d\overline{x}^i\\
d\overline{p}_{ij}
\end{array}
\right)=
S\,\left(\begin{array}{c}
\vartheta_0\\
\vartheta_k\\
d x^k\\
d p_{kl}
\end{array}\right),
\]
\noindent
where $S$ is an analytic function on $J^2(\Ec)$, taking values in the Lie
group $\Gc$ of non-degenerate block matrices of the form
\[
\left(
\begin{array}{cccc}
a               & \tilde{a}^k        & 0               & 0                 \\
\tilde{g}_i     & h_{i}^{k}          & 0               & 0                 \\
\tilde{c}^i     & \tilde{f}^{ik}     & b_{k}^{i}       & r^{ikl}           \\
\tilde{s}_{ij}  & \tilde{w}_{ij}^{k} & \tilde{z}_{ijk} & \tilde{q}_{ij}^{kl}
\end{array}
\right).
\]
\noindent
In this matrix,  $i,j,k,l \in \{1,...,n\}$, $r^{ikl}$ are defined for $k\le l$,
$\tilde{s}_{ij}$, $\tilde{w}_{ij}^{k}$,  and $\tilde{z}_{ijk}$ are defined for
$i\le j$, and  $\tilde{q}_{ij}^{kl}$ are defined for $i\le j$, $k\le l$.

Let us show that $\tilde{a}^k = 0$. Indeed, the exterior (non-closed!) ideal
$\Ic = {\rm span} \{\vartheta_0, \vartheta_i\}$ has the derived ideal
$\delta \Ic = \{\omega\in \Ic \,\vert\, d\omega \in \Ic\}
= {\rm span}\{\vartheta_0\}$. Since $\Delta^{*}\,\overline{\Ic} \subset \Ic$
implies $\Delta^{*}\,(\delta\,\overline{\Ic}) \subset
\delta(\Delta^{*}\,\overline{\Ic}) \subset
\delta\,\Ic$, we obtain $\Delta^{*}\,\overline{\vartheta}_0 = a\,\vartheta_0$.

For convenience in the following computations, we denote by $(B_i^j)$ the
inverse matrix for $(b_i^j)$, so $b_i^j\,B_j^k = \delta_i^k$, by $(H_i^j)$
denote the inverse matrix for $(h_i^j)$, so $h_i^j\,H_j^k = \delta_i^k$,
define $Q_{k^{\prime}l^{\prime}}^{kl}$ by
$Q_{k^{\prime}l^{\prime}}^{kl}\,q^{k^{\prime}l^{\prime}}_{ij} =
\delta_i^k\,\delta_j^l$,
and change the variables on $\Gc$ such that
$g_i= \tilde{g}^i\,a^{-1}$,
$f^{ij} = \tilde{f}^{ik}\,H_{k}^{j}$,
$c^{i}=\tilde{c}^{i}\,a^{-1} - f^{ik}\,g_k$,
$s_{ij} = \tilde{s}_{ij}\,a^{-1} - \tilde{w}_{ij}^{k}\,H_k^m\,g_m -
\tilde{z}_{ijm}\,B_k^m\,c^k$,
$w_{ij}^k=\tilde{w}_{ij}^m\,H_m^k - \tilde{z}_{ijm}\,B_l^m\,f^{lk}$
$z_{ijk} = \tilde{z}_{ijm}\,B_k^m$, and
$q_{ij}^{kl} = \tilde{q}_{ij}^{kl}
- \tilde{z}_{ijm}\,B_{m^{\prime}}^m\,r^{m^{\prime}kl}$.
In accordance with Cartan's method of equivalence, we take the lifted
coframe
\begin{equation}
\left(
\begin{array}{c}
\Theta_0\\
\Theta_i\\
\Xi^i\\
\Sigma_{ij}
\end{array}
\right)=
S\,
\left(\begin{array}{c}
\vartheta_0\\
\vartheta_k\\
dx^k\\
dp_{kl}
\end{array}\right)
=
\left(\begin{array}{l}
a\,\vartheta_0\\
g_i\,\Theta_0 + h_i^k\,\vartheta_k\\
c^i\,\Theta_0+f^{ik}\,\Theta_k+b_k^i\,dx^k + r^{ikl}\,dp_{kl}\\
s_{ij}\,\Theta_0+w_{ij}^{k}\,\Theta_k+z_{ijk}\,\Xi^k + q_{ij}^{kl}\,dp_{kl}
\end{array}\right)
\label{LCF}
\end{equation}
\noindent
on $J^2(\Ec)\times\Gc$. Expressing $du$, $dx^k$, $d p_k$, and $dp_{kl}$ from
(\ref{LCF}) and substituting them to  $d\Theta_0$, we have
\[\fl
d\Theta_0 = da\w \vartheta_0 + a\,d\vartheta_0
= da\,a^{-1}\w \Theta_0 + a\,dx^i\w dp_i
= da\,a^{-1}\w \Theta_0 + a\,dx^i\w \vartheta_i
\]
\[\fl\hspace{25pt}
= \Phi_0^0 \w\Theta_0 + a\,B^i_k\,H^m_i \,\Xi^k \w \Theta_m
+  a\,H^m_i\,R^{ikl}\,\Sigma_{kl}\w\Theta_m
\]
\begin{equation}
\fl\hspace{25pt}
+ a\,H^m_i\,\left(B^i_k\,f^{kj}
+R^{ikl}\,w^j_{kl}\right)\,\Theta_j\w\Theta_m,
\label{dTheta0}
\end{equation}
\noindent
where
\[\fl
\Phi^0_0 = da\,a^{-1} + a\,H^{m^{\prime}}_i\,\left(
B^i_k\left(c^k+R^{ikl}\,s_{kl}\right)\,\Theta_{m^{\prime}}
-g_{m^{\prime}}\,B_k^i\,(\Xi^k-c^k\,\Theta_0-f^{kj}\,\Theta_j)
\right.
\]
\[\fl\hspace{25pt}
\left.
-g_{m^{\prime}}\,R^{ikl}\,(\Sigma_{kl} - s_{kl}\,\Theta_0-w_{kl}^m\,\Theta_m
-z_{klm}\,\Xi^m)
\right)
\]
\noindent
and
$R^{jkl} = - r^{ik^{\prime}l^{\prime}}\,B_i^j\,Q_{k^{\prime}l^{\prime}}^{kl}$.

The multipliers of $\Xi^k \w \Theta_m$, $\Sigma_{kl}\w\Theta_m$, and
$\Theta_j\w\Theta_m$ in (\ref{dTheta0}) are essential  torsion coefficients.
We normalize them by setting $a\,B^i_k\,H^m_i = \delta^m_k$, $R^{ikl} = 0$,
and $f^{kj} = f^{jk}$. Therefore the first normalization is
\begin{equation}
h^k_i = a\,B^k_i, \qquad r^{ikl} = 0, \qquad  f^{kj} = f^{jk}.
\label{normalize1}
\end{equation}
\noindent

Analyzing $d\Theta_i$, $d\Xi^i$, and $d\Sigma_{ij}$ in the same
way, we obtain the following nor\-ma\-li\-za\-ti\-ons:
\begin{equation}
q^{kl}_{ij} = a\,B^k_i\,B^l_j, \qquad s_{ij} = s_{ji}, \qquad
w^k_{ij} = w^k_{ji},\qquad z_{ijk} = z_{jik} = z_{ikj}.
\label{normalize2}
\end{equation}
\noindent
After these reductions  the structure equations for the lifted coframe
have the form
\[\fl
d \Theta_0 = \Phi^0_0 \w \Theta_0 + \Xi^i \w \Theta_i,
\]
\[\fl
d \Theta_i = \Phi^0_i \w \Theta_0 + \Phi^k_i \w \Theta_k
+ \Xi^k \w \Sigma_{ik},
\]
\[\fl
d \Xi^i = \Phi^0_0 \w \Xi^i -\Phi^i_k \w \Xi^k
+\Psi^{i0} \w \Theta_0
+\Psi^{ik} \w \Theta_k,
\]
\[\fl
d \Sigma_{ij} = \Phi^k_i \w \Sigma_{ki} - \Phi^0_0 \w \Sigma_{ij}
+ \Upsilon^0_{ij} \w \Theta_0
+ \Upsilon^k_{ij} \w \Theta_k + \Lambda_{ijk} \w \Xi^k,
\]
\noindent
where the forms $\Phi^0_0$, $\Phi^0_i$, $\Phi^k_i$, $\Psi^{i0}$, $\Psi^{ij}$,
$\Upsilon^0_{ij}$, $\Upsilon^k_{ij}$, and $\Lambda_{ijk}$ are defined by the
following equations:
\[\fl
\Phi^0_0 = da\,a^{-1} - g_k\,\Xi^k+(c^k+f^{km}\,g_m)\,\Theta_k,
\]
\[\fl
\Phi^0_i = d g_i+g_k\,db^k_j\,B^j_i-(g_i\,g_k+s_{ik}+c^j\,z_{ijk})\,\Xi^k
+c^k\,\Sigma_{ik}
\]
\[
+(g_i\,c^k+g_i\,g_m\,f^{mk}-c^j\,w^k_{ij}+f^{mk}\,s_{im})\,\Theta_k,
\]
\[\fl
\Phi^k_i=\delta^k_i\,da\,a^{-1}-d b^k_j\,B^j_i
+(g_i\,\delta^k_j-w^k_{ij}-f^{km}\,z^i_{jm})\,\Xi^j+f^{km}\,\Sigma_{im}
+f^{jm}\,w^k_{ij}\,\Theta_m,
\]
\[\fl
\Psi^{i0} = dc^i+f^{ij}\,\Phi^0_j+c^k\,\Phi^i_k
+(c^i\,f^{mj}\,g_m-c^k\,f^{mj}\,w^i_{kj})\,\Theta_j-c^k\,f^{ij}\,\Sigma_{kj}
\]
\[
+c^k\,(f^{im}\,z^{kmj}+w^i_{kj}-g_k\,\delta^i_j-g_j\,\delta^i_k)\,\Xi^j,
\]
\[\fl
\Psi^{ij} = df^{ij}+(f^{ik}\,\delta^j_m+f^{jk}\,\delta^i_m)\,\Phi^m_k
+(c^i\,\delta^j_k+c^j\,\delta^i_k-f^{ij}\,g_k+f^{im}\,f^{jl}\,z_{klm})\,\Xi^k
\]
\[
+f^{ij}\,(c^k+f^{km}\,g_m)\,\Theta_k - f^{ik}\,f^{jm}\,\Sigma_{km},
\]
\[\fl
\Upsilon^0_{ij} = d s_{ij}-s_{ij}\,da\,a^{-1}+s_{kj}\,db^k_m\,B^m_i
+s_{ik}\,db^k_m\,B^m_j+s_{ij}\,\Phi^0_0+w^k_{ij}\,\Phi^0_k+z_{ijk}\,\Psi^{k0},
\]
\[\fl
\Upsilon^k_{ij} = dw^k_{ij} - w^k_{ij}\,da\,a^{-1}
+(w^k_{il}\,\delta^{m^{\prime}}_j+w^k_{jl}\,\delta^{m^{\prime}}_i)\,db^l_m\,
B^m_{m^{\prime}}
+(s_{ij}\,\delta^k_m+z_{ijl}\,f^{m^{\prime}k}\,w^l_{m^{\prime}m})\,\Xi^m
\]
\[
+w^m_{ij}\,\Phi^k_m
+f^{lk}\,(w^m_{il}\,\delta^{m^{\prime}}_j
+w^m_{jl}\,\delta^{m^{\prime}}_i)\,\Sigma_{m^{\prime}m}
-(c^k+f^{mk}\,g_m)\,\Sigma_{ij},
\]
\[\fl
\Lambda_{ijk} = d z_{ijk} - 2\,z_{ijk}\,da\,a^{-1}
+ z_{ijl}\,db^l_m\,B^m_k
+ z_{ilk}\,db^l_m\,B^m_j
+ z_{ljk}\,db^l_m\,B^m_i
+z_{ijk}\,\Phi^0_0
\]
\[
+z_{ijk}\,g_m\,\Xi^m
+g_i\,\Sigma_{jk}
+g_j\,\Sigma_{ik}
+g_k\,\Sigma_{ij}
-w^l_{ij}\,\Sigma_{lk}
-w^l_{ik}\,\Sigma_{lj}
-w^l_{jk}\,\Sigma_{li}
\]
\[
-f^{lm}\,(
z_{imj}\,\Sigma_{kl}
+z_{imk}\,\Sigma_{jl}
+z_{jmk}\,\Sigma_{il}.
)
\]

Let $\Hc$ be the subgroup of $\Gc$ defined by (\ref{normalize1}) and
(\ref{normalize2}). We shall prove that the restriction of the lifted coframe
(\ref{LCF}) to $J^2(\Ec) \times \Hc$ satisfies  Cartan's test of involutivity,
\cite[def~11.7]{Olver95}. The structure equations remain unchanged under the
following transformation of the modified Maurer - Cartan forms $\Phi^0_0$,
$\Phi^0_i$, $\Phi^k_i$, $\Psi^{i0}$, $\Psi^{ij}$, $\Upsilon^0_{ij}$,
$\Upsilon^k_{ij}$, and $\Lambda_{ijk}$:
\[
\Phi^0_0 \mapsto \Phi^0_0 + K\,\Theta_0,
\]
\[
\Phi^k_i \mapsto \Phi^k_i + L^{kl}_i\,\Theta_l + M^k_i\,\Theta_0,
\]
\[
\Phi^0_i \mapsto \Phi^0_i + M^k_i\,\Theta_k + N_i\,\Theta_0,
\]
\[
\Psi^{ij} \mapsto \Psi^{ij} + P^{ij}\,\Theta_0+S^{ijk}\,\Theta_k
- L^{ij}_k\,\Xi^k,
\]
\[
\Psi^{i0} \mapsto \Psi^{i0} + P^{ij}\,\Theta_j +T^i\,\Theta_0 + K\,\Xi^i
- M^i_k\,\Xi^k,
\]
\[
\Upsilon^{0}_{ij} \mapsto \Upsilon^{0}_{ij} + U_{ij}\,\Theta_0
+ V^k_{ij}\,\Theta_k + W_{ijk}\,\Xi^k + K\,\Sigma_{ij} + M^k_i\,\Sigma_{kj},
\]
\[
\Upsilon^{k}_{ij} \mapsto \Upsilon^{k}_{ij} + X^{kl}_{ij}\,\Theta_l
+ V^k_{ij}\,\Theta_0 + Y^{k}_{ijl}\,\Xi^l + L_i\,\Sigma_{lj},
\]
\[
\Lambda_{ijk} \mapsto \Lambda_{ijk} + Z_{ijkl}\,\Xi^l + Y^l_{ijk}\,\Theta_l
+ W_{ijk}\,\Theta_0,
\]
\noindent
where $K$, $L^{kl}_i$, $M^k_i$, $N_i$, $P^{ij}$, $S^{ijk}$, $T^i$, $U_{ij}$,
$V^k_{ij}$, $W_{ijk}$, $X^{kl}_{ij}$, $Y^k_{ijl}$, and $Z_{ijkl}$ are
arbitrary constants satisfying the following symmetry conditions :
$L^{kl}_i=L^{lk}_i$, $P^{ij}=P^{ji}$,
$S^{ijk}=S^{jik}=S^{ikj}$,
$U_{ij}=U_{ji}$,  $V^k_{ij}=V^k_{ji}$,
$W_{ijk}=W_{jik}=W_{ikj}$,
$X^{kl}_{ij}=X^{kl}_{ji}=X^{lk}_{ij}$, $Y^k_{ijl}=Y^k_{jil}=Y^k_{ilj}$,
and $Z_{ijkl}=Z_{jikl}=Z_{ijlk}=Z_{ikjl}$.  The number of such constants
\[\fl
r^{(1)} = 1 + {{n^2\,(n+1)}\over{2}} +n^2 + n + {{n\,(n+1)}\over{2}}
+ {{n\,(n+1)\,(n+2)}\over{6}}
+ n
+ {{n\,(n+1)}\over{2}}
\]
\[\fl
\hspace{25pt}
+ {{n^2\,(n+1)}\over{2}}
+ {{n\,(n+1)\,(n+2)}\over{6}}
+ {{n^2\,(n+1)^2}\over{4}}
+ {{n^2\,(n+1)\,(n+2)}\over{6}}
\]
\[\fl
\hspace{25pt}
+ {{n\,(n+1)\,(n+2)\,(n+3)}\over{24}} =
{{1}\over{24}}\,(n+1)\,(n+2)\,(11\,n^2+29\,n+12)
\]
\noindent
is the degree of indeterminancy  of the lifted coframe,
\cite[def~11.2]{Olver95}. The reduced characters of this coframe,
\cite[def~11.4]{Olver95},  are easily found
\[
s^{\prime}_i = {{(n+1)\,(n+4)}\over{2}} - i, \qquad i \in \{1, ..., n+1\},
\]
\[
s^{\prime}_{n+1+j} = {{(n+1-j)\,(n+2-j)}\over{2}}, \qquad j \in \{1, ..., n\}.
\]
\noindent
A simple calculation shows that
\[
r^{(1)} =
s^{\prime}_1
+ 2\,s^{\prime}_2
+ 3\,s^{\prime}_3
+ ...
+ (2\,n+1)\,s^{\prime}_{2\,n+1}.
\]
\noindent
So the Cartan test is satisfied, and the lifted coframe is involutive.

It is easy to directly verify that a transformation
$\hat{\Delta} : J^2(\Ec) \times \Hc \rightarrow J^2(\Ec) \times \Hc$
satisfies the conditions
\begin{equation}
\hat{\Delta}^{*}\, \overline{\Theta}_0 = \Theta_0,
\qquad
\hat{\Delta}^{*}\, \overline{\Theta}_i = \Theta_i,
\qquad
\hat{\Delta}^{*}\, \overline{\Xi}^i = \Xi^i,
\qquad
\hat{\Delta}^{*}\, \overline{\Sigma}_{ij} = \Sigma^{ij}
\label{def_cond}
\end{equation}
\noindent
if and only if it is projectable on $J^2(\Ec)$, and its projection
$\Delta : J^2(\Ec) \rightarrow J^2(\Ec)$ is a contact transformation.

Since (\ref{def_cond}) imply
$\hat{\Delta}^{*}\,d\overline{\Theta}_0 = d\Theta_0$,
$\hat{\Delta}^{*}\,d\overline{\Theta}_i = d\Theta_i$,
$\hat{\Delta}^{*}\,d\overline{\Xi}^i = d\Xi^i$,
and
$\hat{\Delta}^{*}\,d\overline{\Sigma}_{ij} = d\Sigma_{ij}$,
we have
\[\fl
\hat{\Delta}^{*} \left(\overline{\Phi}^0_0 \w \overline{\Theta}_0
+ \overline{\Xi}^i \w \overline{\Theta}_i\right)
=
\left(\hat{\Delta}^{*} \overline{\Phi}^0_0\right) \w \Theta_0
+ \Xi^i \w \Theta_i
=
\Phi^0_0 \w \Theta_0 + \Xi^i \w \Theta_i,
\]
\[\fl
\hat{\Delta}^{*} \left(\overline{\Phi}^0_i \w \overline{\Theta}_0
+ \overline{\Phi}^k_i \w \overline{\Theta}_k
+ \overline{\Xi}^k \w \overline{\Sigma}_{ik}\right)
=
\hat{\Delta}^{*} \left(\overline{\Phi}^0_i\right) \w \Theta_0
+ \hat{\Delta}^{*} \left(\overline{\Phi}^k_i\right) \w \Theta_k
+ \Xi^k \w \Sigma_{ik}
\]
\[\fl\hspace{25pt}
=
\Phi^0_i \w \Theta_0
+ \Phi^k_i \w \Theta_k
+ \Xi^k \w \Sigma_{ik},
\]
\[\fl
\hat{\Delta}^{*}\left(\overline{\Phi}^0_0 \w \overline{\Xi}^i
- \overline{\Phi}^i_k \w \overline{\Xi}^k
+\overline{\Psi}^{i0} \w \overline{\Theta}_0
+\overline{\Psi}^{ik} \w \overline{\Theta}_k\right)
\]
\[\fl\hspace{25pt}
=
\hat{\Delta}^{*}\left(\overline{\Phi}^0_0\right) \w \Xi^i
-\hat{\Delta}^{*}\left(\overline{\Phi}^i_k\right) \w \Xi^k
+\hat{\Delta}^{*}\left(\overline{\Psi}^{i0}\right) \w \Theta_0
+\hat{\Delta}^{*}\left(\overline{\Psi}^{ik}\right) \w \Theta_k
\]
\[\fl\hspace{25pt}
=
\Phi^0_0 \w \Xi^i -\Phi^i_k \w \Xi^k
+\Psi^{i0} \w \Theta_0
+\Psi^{ik} \w \Theta_k,
\]
\[\fl
\hat{\Delta}^{*}\left(
\overline{\Phi}^k_i \w \overline{\Sigma}_{ki}
- \overline{\Phi}^0_0 \w \overline{\Sigma}_{ij}
+ \overline{\Upsilon}^0_{ij} \w \overline{\Theta}_0
+ \overline{\Upsilon}^k_{ij} \w \overline{\Theta}_k
+ \overline{\Lambda}_{ijk} \w \overline{\Xi}^k
\right)
\]
\[\fl\hspace{25pt}
=
\hat{\Delta}^{*}\left(\overline{\Phi}^k_i\right) \w \Sigma_{ki}
- \hat{\Delta}^{*}\left(\overline{\Phi}^0_0\right) \w \Sigma_{ij}
+ \hat{\Delta}^{*}\left(\overline{\Upsilon}^0_{ij}\right) \w \Theta_0
+ \hat{\Delta}^{*}\left(\overline{\Upsilon}^k_{ij}\right) \w \Theta_k
\]
\[\fl\hspace{25pt}
+ \hat{\Delta}^{*}\left(\overline{\Lambda}_{ijk}\right) \w \Xi^k
=
\Upsilon^0_{ij} \w \Theta_0
- \Phi^0_0 \w \Sigma_{ij} + \Phi^k_i \w \Sigma_{ki}
+ \Upsilon^k_{ij} \w \Theta_k + \Lambda_{ijk} \w \Xi^k.
\]
\noindent
Therefore,
\[
\hat{\Delta}^{*}\left(\overline{\Phi}^0_0\right) = \Phi^0_0 + K\,\Theta_0,
\]
\[
\hat{\Delta}^{*}\left(\overline{\Phi}^k_i\right) = \Phi^k_i
+ L^{kl}_i\,\Theta_l + M^k_i\,\Theta_0,
\]
\[
\hat{\Delta}^{*}\left(\overline{\Phi}^0_i\right) = \Phi^0_i + M^k_i\,\Theta_k
+ N_i\,\Theta_0,
\]
\begin{equation}
\hat{\Delta}^{*}\left(\overline{\Psi}^{ij}\right) = \Psi^{ij}
+ P^{ij}\,\Theta_0 + S^{ijk}\,\Theta_k - L^{ij}_k\,\Xi^k,
\label{Gamma_rules}
\end{equation}
\[
\hat{\Delta}^{*}\left(\overline{\Psi}^{i0}\right) = \Psi^{i0}
+ P^{ij}\,\Theta_j + T^i\,\Theta_0 + K\,\Xi^i - M^i_k\,\Xi^k,
\]
\[
\hat{\Delta}^{*}\left(\overline{\Upsilon}^{0}_{ij}\right) =  \Upsilon^{0}_{ij}
+ U_{ij}\,\Theta_0 + V^k_{ij}\,\Theta_k + W_{ijk}\,\Xi^k + K\,\Sigma_{ij}
+ M^k_i\,\Sigma_{kj},
\]
\[
\hat{\Delta}^{*}\left(\overline{\Upsilon}^{k}_{ij}\right) = \Upsilon^{k}_{ij}
+ X^{kl}_{ij}\,\Theta_l + V^k_{ij}\,\Theta_0 + Y^{k}_{ijl}\,\Xi^l
+ L_i\,\Sigma_{lj},
\]
\[
\hat{\Delta}^{*}\left(\overline{\Lambda}_{ijk}\right) = \Lambda_{ijk}
+ Z_{ijkl}\,\Xi^l + Y^l_{ijk}\,\Theta_l
+ W_{ijk}\,\Theta_0,
\]
with some functions $K$, $L^{kl}_i$, $M^k_i$, $N_i$, $P^{ij}$, $S^{ijk}$,
$T^i$, $U_{ij}$, $V^k_{ij}$, $W_{ijk}$, $X^{kl}_{ij}$, $Y^k_{ijl}$, and
$Z_{ijkl}$ on $J^2(\Ec) \times \Hc$.

\section{Contact symmetries of differential equations}

Suppose $\Rc$ is a second-order differential equation in one dependent and
$n$ independent variables. We consider $\Rc$ as a sub-bundle in $J^2(\Ec)$.
Let $Cont(\Rc)$ be the group of contact symmetries for $\Rc$. It consists of
all the contact transformations on $J^2(\Ec)$ mapping $\Rc$ to itself.
The moving coframe method, \cite{FO,FO2}, is applicable to find invariant
1-forms characterizing $Cont(\Rc)$ is the same way, as the restriction of
the lifted coframe (\ref{LCF}) to $J^2(\Ec)\times\Hc$ characterizes
$Cont(J^2(\Ec))$. We briefly outline this approach.

Let $\iota : \Rc \rightarrow J^2(\Ec)$ be an embedding. The invariant
1-forms of $Cont(\Rc)$ are re\-stric\-ti\-ons of the coframe (\ref{LCF}),
(\ref{normalize1}), (\ref{normalize2}) to $\Rc$:
$\theta_0 = \iota^{*} \Theta_0$, $\theta_i= \iota^{*}\Theta_i$,
$\xi^i = \iota^{*}\Xi^i$, and $\sigma_{ij}=\iota^{*}\Sigma_{ij}$
(for brevity we identify the map
$\iota \times id : \Rc\times \Hc \rightarrow J^2(\Ec)\times \Hc$ with
$\iota : \Rc \rightarrow J^2(\Ec)$). The forms $\theta_0$, $\theta_i$,
$\xi^i$, and $\sigma_{ij}$ have some linear dependencies, i.e., there exists a
non-trivial set of functions $E^0$, $E^i$, $F_i$, and $G^{ij}$ on
$\Rc\times \Hc$ such that
$E^0\,\theta_0 + E^i\,\theta_i + F_i\,\xi^i+ G^{ij}\,\sigma_{ij} \equiv 0$.
These functions are lifted invariants of $Cont(\Rc)$. Setting them equal to
some constants allows us to specify some parameters $a$, $b^k_i$, $c_i$,
$g_i$, $f^{ij}$, $s_{ij}$, $w^k_{ij}$, and $z_{ijk}$ of the group $\Hc$
as functions of the coordinates on $\Rc$ and the other group parameters.

After these normalizations, some restrictions of the modified
Maurer - Cartan forms $\phi^0_0=\iota^{*}\Phi^0_0$,
$\phi^k_i=\iota^{*}\Phi^k_i$, $\phi^0_i=\iota^{*}\Phi^0_i$,
$\psi^{ij}=\iota^{*}\Psi^{ij}$, $\psi^{i0}=\iota^{*}\Psi^{i0}$,
$\upsilon^{0}_{ij}=\iota^{*}\Upsilon^{0}_{ij}$,
$\upsilon^{k}_{ij}=\iota^{*}\Upsilon^{k}_{ij}$, and
$\lambda_{ijk}=\iota^{*}\Lambda_{ijk}$, or some their linear
combinations, become semi-basic, i.e., they do not include the
differentials of the parameters of $\Hc$. From
(\ref{Gamma_rules}), we have the following statements: (i) if
$\phi^0_0$ is semi-basic, then its coefficients at $\theta_k$,
$\xi^k$, and $\sigma_{kl}$ are lifted invariants of $Cont(\Rc)$;
(ii) if $\phi^0_i$ or $\phi^k_i$ are semi-basic, then their
coefficients at $\xi^k$ and $\sigma_{kl}$ are lifted invariants of
$Cont(\Rc)$; (iii) if $\psi^{i0}$, $\psi^{ij}$, or $\lambda_{ijk}$
are semi-basic, then their coefficients  at $\sigma_{kl}$ are
lifted invariants of $Cont(\Rc)$. Setting these invariants equal
to some constants, we get specifications of some more parameters
of $\Hc$ as functions of the coordinates on $\Rc$ and the other
group parameters.

More lifted invariants can appear as essential torsion coefficients in the
reduced structure equations
\[
d \theta_0 = \phi^0_0 \w \theta_0 + \xi^i \w \theta_i
\]
\[
d \theta_i =
\phi^0_i \w \theta_0 + \phi^k_i \w \theta_k + \xi^k \w \sigma_{ik}
\]
\[
d \xi^i = \phi^0_0 \w \xi^i -\phi^i_k \w \xi^k +\psi^{i0} \w \theta_0
+\psi^{ik} \w \theta_k
\]
\[
d \sigma_{ij} = \phi^k_i \w \sigma_{ki} - \phi^0_0 \w \sigma_{ij}
+ \upsilon^0_{ij} \w \theta_0 + \upsilon^k_{ij} \w \theta_k
+ \lambda_{ijk} \w \xi^k.
\]
\noindent
After normalizing these invariants and repeating the process, two outputs are
possible. In the first case, the reduced lifted coframe appears to be
involutive. Then this coframe is the desired set of defining forms for
$Cont(\Rc)$. In the second case, when the reduced lifted coframe does not
satisfy Cartan's test, we should use the procedure of prolongation,
\cite[ch~12]{Olver95}.

\section{Structure and invariants of symmetry groups for linear parabolic
equations}

We apply the method described in the previous section to the class of linear
parabolic equations (\ref{para}). Denote $x^1=t$, $x^2=x$, $p_1=u_t$,
$p_2=u_x$, $p_{11}=u_{tt}$, $p_{12}=u_{tx}$, and $p_{22}=u_{xx}$.
The coordinates on $\Rc$ are $\{t, x, u, u_t, u_x, u_{tt}, u_{tx}\}$,
and the embedding $\iota : \Rc \rightarrow J^2(\Ec)$ is defined by
(\ref{para}). Computing the linear dependence conditions for the
reduced forms $\theta_0$, $\theta_i$, $\xi^i$, and $\sigma_{ij}$ by means
of {\sc MAPLE}, we express the group parameters $b^1_2$, $z_{122}$,
$z_{222}$, $w^1_{22}$, $w^2_{22}$, and $s_{22}$
as functions of the coordinates on $\Rc$ and the other parameters of the
group $\Hc$. Particularly, since
\[
\sigma_{22} \equiv -(b^1_2)^2(b^2_2)^{-2}\,\sigma_{11}
-2 b^1_2 (b^2_2)^{-1}\,\sigma_{12} \quad
(\rm{mod} \,\,\theta_0, \theta_1, \theta_2, \xi^1, \xi^2),
\]
\noindent and without loss of generality $b^1_1 \not = 0$,
$b^2_2\not=0$, we take $b^1_2 = 0$. After that, we have
\[\fl
\sigma_{22} \equiv \left(z^1_{22}+a\,
\left(b^2_2\left(T\,u_{tt}+X\,u_{tx}+(U+T_t)\,u_t+X_t\,u_x+U_t\,u\right)
\right.
\right.
\]
\[\fl\hspace{15pt}
\left.
\left.
-b^2_1\left(T\,u_{tx}+(T\,X+T_x)\,u_t+(X^2+U+X_x)\,u_x+(U_x+X\,U)\,u\right)
\right)
\,(b^1_1)^{-1}(b^2_2)^{-3}
\right)\,\xi^1
\]
\[\fl\hspace{15pt}
+
\left(
z^2_{22}+
a\,\left(
T\,u_{tx}+(T\,X+T_x)\,u_t+(X^2+U+X_x)\,u_x+(U_x+X\,U)\,u
\right)\,(b^2_2)^3
\right)\,\xi^2\,
\]
\[\fl\hspace{15pt}
(\rm{mod}\,\, \theta_0, \theta_1, \theta_2).
\]
\noindent
Then we take
\[\fl
z^1_{22} = - a\,
\left(b^2_2\left(T\,u_{tt}+X\,u_{tx}+(U+T_t)\,u_t+X_t\,u_x+U_t\,u\right)
\right.
\]
\[\fl\hspace{15pt}
\left.
-b^2_1\left(T\,u_{tx}+(T\,X+T_x)\,u_t+(X^2+U+X_x)\,u_x+(U_x+X\,U)\,u\right)
\right)
\,(b^1_1)^{-1}(b^2_2)^{-3},
\]
\[\fl
z^2_{22} =-
a\,\left(
T\,u_{tx}+(T\,X+T_x)\,u_t+(X^2+U+X_x)\,u_x+(U_x+X\,U)\,u
\right)\,(b^2_2)^3.
\]
\noindent
After that, setting the coefficients of $\sigma_{22}$ at $\theta_1$,
$\theta_2$, and  $\theta_0$ equal to $0$, we find $w^1_{22}$, $w^2_{22}$,
and $s_{22}$ as the functions of the coordinates on $\Rc$ and the other
parameters of $\Hc$. These expressions are too long to be written out in
full here.

Now the form $\phi^1_2$ is semi-basic. We have
\[
\phi^1_2 \equiv f^{11}\,\sigma_{12} +b^1_1\,T\,(b^2_2)^{-2}\,\xi^2
\quad (\rm{mod}\,\, \theta_0, \theta_1, \theta_2, \xi^1, \sigma_{11}),
\]
\noindent
therefore we take $f^{11} = 0$, $b^1_1= (b^2_2)^2\,T^{-1}$. After that, setting
the coefficient of $\phi^1_2$ at $\xi^1$ equal to $0$, we find $w^1_{12}$.

Then the linear combination $2\,\phi^2_2-\phi^1_1-\phi^0_0$ becomes semi-basic.
Since
\[\fl
2\,\phi^2_2-\phi^1_1-\phi^0_0 \equiv
f^{12}\,\sigma_{12}
+\left(4\,g_2
+
\left(
2\,T^2\,b^2_1+(2\,T\,X-T_x)\,b^2_2
\right)\,(b^2_2)^{-2}\,T^{-1}\right)
\,\xi^2
\]
\[\fl\hspace{25pt}
(\rm{mod}\,\, \theta_0, \theta_1, \theta_2, \xi^1, \sigma_{11}),
\]
\noindent
we take $f^{12} = 0$,
$g_2 = - \left(2\,T^2\,b^2_1+(2\,T\,X-T_x)\,b^2_2\right)/(4\,(b^2_2)^2\,T)$.
Setting the co\-ef\-fi\-ci\-ent of $2\,\phi^2_2-\phi^1_1-\phi^0_0$ at $\xi^1$
equal to $0$, we find $w^2_{12}$.

Since for the semi-basic linear combination $2\,\phi^0_2-\phi^2_1$
we have $2\,\phi^0_2-\phi^2_1 \equiv (2\,c^1-f^{22})\,\sigma_{12}
\quad (\rm{mod}\,\, \theta_0, \theta_1, \theta_2, \xi^1, \xi^2,
\sigma_{11})$, the normalization $c^{1} = f^{22}/2$ is possible. Setting the
co\-ef\-fi\-ci\-ent of $2\,\phi^0_2-\phi^2_1$ at $\xi^1$ and $\xi^2$ equal to
$0$, we find $s_{12}$ and $g_1$.

After that, we obtain the following reduced structure equations
\[
d \theta_0 = \alpha_1\w\theta_0 +\xi^1\w\theta_1+\xi^2\w\theta_2,
\]
\[
d \theta_2 = \alpha_1\w\theta_2-\case12\,\alpha_2\w\theta_2+\alpha_3\w\theta_0
+\xi^1\w\sigma_{12}+\xi^2\w\theta_1+\case14\,f^{22}\,\theta_1\w\theta_2,
\]
\[
d \xi^1 = \alpha_2\w\xi^1 + \alpha_4\w\theta_0+\case12\,f^{22}\,\xi^2\w\theta_2,
\]
\noindent
where $\alpha_1$, $\alpha_2$, $\alpha_3$, and $\alpha_4$ are 1-forms on
$J^2(\Ec)\times\Hc$ depending on differentials of the parameters of $\Hc$.
We normalize the essential torsion coefficient $f^{22}$ in these
equations by setting $f^{22}=0$.
Then, there are the following structure equations
\[\fl
d \theta_0 = \alpha_1\w\theta_0 +\xi^1\w\theta_1+\xi^2\w\theta_2,
\]
\[\fl
d \theta_1 = \alpha_1\w\theta_1-\alpha_2\w\theta_1+2\,\alpha_3\w\theta_2
+\alpha_4\w\theta_0+\xi^1\w\sigma_{11}+\xi^2\w\sigma_{12}
-c^2\,\theta_1\w\theta_2,
\]
\[\fl
d \theta_2 = \alpha_1\w\theta_2-\case12\,\alpha_2\w\theta_2+\alpha_3\w\theta_0
+\xi^1\w\sigma_{12}+\xi^2\w\theta_1,
\]
\[\fl
d \xi^1 = \alpha_2\w\xi^1
\]
\noindent
(the forms $\alpha_i$ can change after the normalizations).
The structure equations have the essential torsion coefficient $c^2$, therefore
we normalize $c^2=0$. After that, we set the coefficients of $d\sigma_{12}$
at $\theta_0\w\xi^2$ and $\theta_2\w\xi^2$ equal to $0$ and express
$w^2_{11}$ and $s_{11}$ as functions of the coordinates on $\Rc$
and the remaining parameters of $\Hc$. The formulas for $w^2_{11}$ and
$s_{11}$ are too long to be written out in full here.
Then we get
\[\fl
d \sigma_{11} = \alpha_1\w\sigma_{11}-2\,\alpha_2\w\sigma_{11}
+4\,\alpha_3\w\sigma_{12}+6\,\alpha_4\w\theta_1+\alpha_5\w\xi^2
+\alpha_6\w\xi^1
\]
\[\fl\hspace{25pt}
+I^5\,(b^2_2)^{-5}\,\theta_0\w\xi^2,
\]
\[\fl
d \tau_{12} = \alpha_1\w\sigma_{12}-\case32\,\alpha_2\w\sigma_{12}
+3\,\alpha_3\w\theta_1+3\,\alpha_4\w\theta_2+\alpha_5\w\xi^1
+\xi^2\w\sigma_{11},
\]
\noindent
where
$I^5  = - \case1{16} \,\lambda\,T^5$,
\noindent
$\lambda$ is given by (\ref{JM_lambda}), and all the essential torsion
coefficients in the other structure equations are constants.

There are two possibilities now: $I = 0$ or $I \not= 0$.
By $\Pc_1$ we denote the subclass of all equations (\ref{para}) such that
$I\not =0$. For an equation from $\Pc_1$ all the essential torsion
coefficients in the reduced structure equations are constants, but the lifted
coframe $\theta_0$, $\theta_1$, $\theta_2$, $\xi^1$, $\xi^2$, $\sigma_{11}$,
and $\sigma_{12}$ is not involutive yet. Therefore we use the procedure of
prolongation, \cite[Ch 12]{Olver95}, and obtain the structure equations
\[\fl
d\theta_0=\alpha_1\w\theta_0+\xi^1\w\theta_1+\xi^2\w\theta_2,
\]
\[\fl
d\theta_1=
\alpha_1\w\theta_1
-
\alpha_2\w\theta_1
+
2\,
\alpha_3\w\theta_2
+
\alpha_4\w\theta_0
+
\xi^2\w\sigma_{12}+
\xi^1\w\sigma_{11},
\]
\[\fl
d\theta_2=
\alpha_1\w\theta_2
-\case12\,\alpha_2\w\theta_2
+\alpha_3\w\theta_0
+
\xi^1\w\sigma_{12}
-
\theta_1\w\xi^2,
\]
\[\fl
d\xi^1
=\alpha_2\w\xi^1,
\]
\[\fl
d\xi^2=
-2\,
\alpha_3\w\xi^1+\case12\,\alpha_2\w\xi^2,
\]
\[\fl
d\sigma_{11}=
\alpha_1\w\sigma_{11}
-2\,\alpha_2\w\sigma_{11}
+4\,\alpha_3\w\sigma_{12}
+6\,\alpha_4\w\theta_1
+\alpha_5\w\xi^2
+\alpha_6\w\xi^1,
\]
\[\fl
d\sigma_{12}=
\alpha_1\w\sigma_{12}
-\case32\,\alpha_2\w\sigma_{12}
+3\,\alpha_3\w\theta_1
+3\,\alpha_4\w\theta_2
+\alpha_5\w\xi^1
+\xi^2\w\sigma_{11},
\]
\[\fl
d\alpha_1=
-\alpha_3\w\xi^2
-\alpha_4\w\xi^1,
\]
\[\fl
d\alpha_2=
4\,\alpha_4\w\xi^1,
\]
\[\fl
d\alpha_3=
-\alpha_4\w\xi^2
-\case12\,\alpha_2\w\alpha_3,
\]
\[\fl
d\alpha_4=-\alpha_2\w\alpha_4,
\]
\[\fl
d\alpha_5=
\pi_1\w\xi^1
+\alpha_1\w\alpha_5
-\case52\,\alpha_2\w\alpha_5
-5\,\alpha_3\w\sigma_{11}
-10\,\alpha_4\w\sigma_{12}
-\alpha_6\w\xi^2,
\]
\[\fl
d\alpha_6=
\pi_1\w\xi^2
+\pi_2\w\xi^1
+\alpha_1\w\alpha_6
-3\,\alpha_2\w\alpha_6
+6\,\alpha_3\w\alpha_5
-15\,\alpha_4\w\sigma_{11},
\]
\noindent
where $\alpha_1 ,... ,\alpha_6$, $\pi_1$, and $\pi_2$  are 1-forms on
$\Rc\times\Hc$ (they are too long to be written explicitly). From these
structure equations, it follows that the only non-zero reduced character of
the lifted coframe $\theta_0$, $\theta_1$, $\theta_2$, $\xi^1$, $\xi^2$,
$\sigma_{11}$, $\sigma_{12}$, $\alpha_1$, $\alpha_2$, ..., $\alpha_6$
is $s^{\prime}_1 =2$, while the degree of indeterminancy is $r^{(1)} = 2$.
So the Cartan test is satisfied, and the lifted coframe is involutive.

The same calculations show that the symmetry group of the linear heat
equation
\begin{equation}
u_{xx} = u_t
\label{heat}
\end{equation}
has the identical structure equations, but with the different lifted coframe.
All the essential torsion coefficients in the structure equations are
constants. Thus, applying Theorem 15.12 of \cite{Olver95}, we have

\vskip 5 pt
\noindent
{\bf Theorem 1.} (\cite{JM}, Theorem 3.2)
{\it
The linear parabolic equation (\ref{para}) is equivalent to the linear
heat equation (\ref{heat}) under a contact transformation if and only if
it belongs to} $\Pc_1$, {\it i.e., iff} $I = 0$.

\vskip 5 pt Numerous examples of equations (\ref{para}) reducible
to the linear heat equation are given in \cite{JM}, \cite{Stohny}.

\vskip 10 pt

Now we return to the case $I\not=0$. Then we can take $b^2_2 = I$. Setting the
essential torsion coefficient in the structure equation for $d\theta_2$ at
$\theta_2\w\xi^2$ equal to $0$ and expressing $w^1_{11}$, we get the following
structure equations
\[\fl
d\theta_0=\alpha_1\w\theta_0+\xi^1\w\theta_1+\xi^2\w\theta_2,
\]
\[\fl
d\theta_1 = \alpha_1\w\theta_1+2\,\alpha_2\w\theta_2
-\case12\,J_1\,\alpha_2\w\theta_0 + Z\,\theta_0\w\xi^1
-\left(b^2_1\,J_{1x} -I\,J_{1t}\right)/(4\,I^3)\,\theta_0\w\xi^1
\]
\[\fl\hspace{20pt}
-J_1\,\xi^2\w\theta_1+\xi^1\w\sigma_{11}+\xi^2\w\sigma_{12},
\]
\[\fl
d\theta_2 = \alpha_1\w\theta_2+\alpha_2\w\theta_0+\xi^2\w\theta_1
-\case12\,J_1\,\xi^2\w\theta_2+\xi^1\w\sigma_{12},
\]
\[\fl
d\xi^1 = -J_1\,\xi^1\w\xi^2,
\]
\[\fl
d\xi^2 = -2\,\alpha_2\w\xi^1,
\]
\noindent
where
\[
J_1 = \left(2\,T\,I_x-I\,T_x\right)\,T^{-1}\,I^{-2},
\]
\noindent
and $Z$ is a function of $T$, $X$, $U$, $I$, $J_1$, their derivatives w.r.t.
$t$, $x$, and $b^2_1$. Recall that the forms $\alpha_1$, $\alpha_2$ are not
necessary the same as in the previous structure equations.

Consider the subclass $\Pc_2$ of all equations (\ref{para}) such that
$I\not = 0$, $J_{1x}\not = 0$. This subclass is not empty, e.g., the equation
$u_{xx} = u_t + x^4\,u$ belongs to $\Pc_2$. For an equation from $\Pc_2$, we
normalize the coefficient in the structure equation for $d\theta_1$ at
$\theta_0\w\xi^1$ by setting $b^2_1 = -I\,J_{1t}\,J_{1x}^{-1}$. Then, after
a prolongation, we obtain the following structure equations
\[\fl
d\theta_0=\alpha_1\w\theta_0
+\xi^1\w\theta_1
+\xi^2\w\theta_2,
\]
\[\fl
d\theta_1=
\alpha_1\w\theta_1
+J_3\,\theta_0\w\xi^1
-\case14\,J_1\,J_2\,\theta_0\w\xi^2
+J_1\,\theta_1\w\xi^2
+J_4\,\theta_2\w\xi^1
\]
\[\fl\hspace{25pt}
+J_2\,\theta_2\w\xi^2
+\xi^1\w\sigma_{11}
+\xi^2\w\sigma_{12},
\]
\[\fl
d\theta_2=
\alpha_1\w\theta_2
+\case12\,J_4\,\theta_0\w\xi^1
+\case12\,J_2\,\theta_0\w\xi^2
+\case12\,J_1\,\theta_2\w\xi^2
+\xi^1\w\sigma_{12}
-\theta_1\w\xi^2,
\]
\[\fl
d\xi^1=-J_1\,\xi^1\w\xi^2,
\]
\begin{equation}
\fl
d\xi^2=-J_2\,\xi^1\w\xi^2,
\label{SE_B}
\end{equation}
\[\fl
d\sigma_{11}=
\alpha_1\w\sigma_{11}
+\alpha_2\w\xi^1
+\alpha_3\w\xi^2,
\]
\[\fl
d\sigma_{12}=
\alpha_1\w\sigma_{12}
+\alpha_3\w\xi^1
-\theta_0\w\xi^1
+\case32\, (J_4+J_1\,J_2 )\,\theta_1\w\xi^1
+\case32\,J_2\,\theta_1\w\xi^2
\]
\[\fl\hspace{25pt}
+3\,J_3\,\theta_2\w\xi^1
-\case34\,J_1\,J_2\,\theta_2\w\xi^2
+2\,J_1\,\xi^1\w\sigma_{11}
+2\,J_2\,\xi^1\w\sigma_{12}
+\xi^2\w\sigma_{11}
\]
\[\fl\hspace{25pt}
-\case32\,J_1\,\xi^2\w\sigma_{12}
\]
\[\fl
d\alpha_1
=\case14\,(2\,J_4+J_1\,J_2)\,
\xi^1\w\xi^2,
\]
\[\fl
d\alpha_2=
\pi_1\w\xi^1
+\pi_2\w\xi^2
+\alpha_1\w\alpha_2
+J_1\,\alpha_2\w\xi^2
+J_2\,\alpha_3\w\xi^2
-\case14\,(2\,J_4+J_1J_2)\,\xi^2\w\sigma_{11}
\]
\[\fl
d\alpha_3=
\pi_2\w\xi^1+\alpha_1\w\alpha_3-\alpha_2\w\xi^2
+\case92\,J_1\,\alpha_3\,\w\xi^2
-(\case52\,J_1-\case38\,{J_1}^{2}{J_2}^{2})\,\theta_0\w\xi^2
\]
\[\fl\hspace{25pt}
+\case34\,(2\,\Dc_2(J_4)+2\,J_1\,\Dc_2(J_2)+2\,J_2\,\Dc_2(J_1)
+6\,{J_2}^{2}+3\,J_4\,J_1+3\,{J_1}^{2}\,J_2+4\,J_3
\]
\[\fl\hspace{25pt}
-2\,\Dc_1(J_2))\,\theta_1\w\xi^2
+(3\,\Dc_2(J_3)-\case{15}{4}\,J_1\,{J_2}^{2}+6\,J_3\,J_1-1
+\case34\,J_1\Dc_1(J_2))\,\theta_2\w\xi^2
\]
\[\fl\hspace{25pt}
+(\case92\,J_2+5\,{J_1}^{2}+2\,\Dc_2(J_1))\,\xi^2\w\sigma_{11}
+2\,(\Dc_2(J_2)-J_1J_2-2\,J_4)\,\xi^2\w\sigma_{12},
\]
\noindent
where $\alpha_1$, $\alpha_2$, $\alpha_3$, $\pi_1$, and $\pi_2$ are 1-forms
on $\Rc\times\Hc$. The functions $J_2$, $J_3$, and $J_4$ are defined as
follows:
\[\fl
J_2=
1/2\,\left(
2\,T\,I\,J_{{1x}}\,J_{{1tx}}
-2\,T\,I_{{t}}\,J_{{1x}}^{2}
-2\,T\,I\,J_{{1t}}\,J_{{1xx}}
+T\,I^{2}\,J_1\,J_{{1t}}\,J_{{1x}}\,
\right.
\]
\[\fl\hspace{25pt}
\left.
+I\,T_x\,J_{{1t}}\,
J_{{1x}}\right)\,
\,J_{{1x}}^{-2}\,I^{-3},
\]
\[\fl
J_3=1/32\,(-135\,I^2\,T_x^4\,J_{1x}^2-32\,T^6\,I_{t}^2\,J_{1x}^2
+16\,J_{1x}^2\,T^{3}\,I^2\,T_{xx}\,T_t+16\,J_{1x}^2\,T^{5}\,I^2\,X_{tx}
\]
\[\fl\hspace{25pt}
+216\,I^2\,T_x^2\,T\,T_{xx}\,J_{1x}^2+32\,T^4\,I^2\,U_{xx}\,J_{1x}^2
+16\,T^3\,I^2\,X_x\,T_{xx}\,J_{1x}^2+8\,T^3\,I^2\,T_{xxxx}\,J_{1x}^2
\]
\[\fl\hspace{25pt}
-32\,I^2\,U\,T^3\,T_{xx}\,J_{1x}^2-36\,T^2\,I^2T_{xx}^2\,J_{1x}^2
-16\,T^4\,I^2\,T_{txx}\,J_{1x}^2+16\,J_{1x}^2\,T^6\,I\,I_{tt}
\]
\[\fl\hspace{25pt}
+16\,T^4\,I^2\,X_{x}^2\,J_{1x}^2
+8\,T^6\,I_{t}\,J_{1x}\,I^2\,J_{1t}\,J_1
-40\,I^2\,T_x^2\,T^2X_{x}J_{1x}^2+8\,T^6I^3J_{1t}J_{1tx}J_1
\]
\[\fl\hspace{25pt}
-8\,J_{1t}^2T^6J_{1x}I^3-8\,J_{1x}T^6I^3J_{1tt}J_1
-8\,T^5I^3T_tJ_{1x}J_{1t}J_1-8\,T^3I^2X^2T_{xx}J_{1x}^2
\]
\[\fl\hspace{25pt}
+8\,J_{1t}T^5I^5J_1J_{1x}J_2-4\,J_{1t}^2T^6I^4J_1^2
+16\,I^2T^4X_{xx}XJ_{1x}^2+40\,T^3I^2T_xT_{tx}J_{1x}^2
\]
\[\fl\hspace{25pt}
+20\,I^2T_x^2T^2X^2J_{1x}^2-8\,T^4I^2X_tT_xJ_{1x}^2
-40\,I^2T^3XX_{x}T_xJ_{1x}^2-40\,I^2T_x^2T^2T_tJ_{1x}^2
\]
\[\fl\hspace{25pt}
-56\,T^2I^2T_{xxx}T_xJ_{1x}^2-16\,T^4I^2X_{xxx}J_{1x}^2
+16\,T^5IT_tJ_{1x}^2I_{t}+40\,I^2T^3X_{xx}T_xJ_{1x}^2
\]
\[\fl\hspace{25pt}
+80\,I^2T_x^2UT^2J_{1x}^2-80\,I^2T^3U_xT_xJ_{1x}^2)
\,
T^{-4}
J_{1x}^{-2}
I^{-6},
\]
\[\fl
J_4=1/8\,(-8\,J_{1t}^2T^6I^3-32\,T^6I_{t}^2J_{1x}
-135\,I^2T_x^4J_{1x}+16\,T^{5}IT_tJ_{1x}I_{t}
\]
\[\fl\hspace{25pt}
-8\,T^4I^2X_tT_xJ_{1x}+20\,I^2T_x^2T^2X^2J_{1x}
+40\,I^2T^3X_{xx}T_xJ_{1x}
-8\,I^3J_{1x}T^4X_{xx}J_1
\]
\[\fl\hspace{25pt}
-80\,I^2T^3U_xT_xJ_{1x}-16\,T^4I^2X_{xxx}J_{1x}
-40\,I^2T^3XX_xT_xJ_{1x}-8\,T^3I^2X^2T_{xx}J_{1x}
\]
\[\fl\hspace{25pt}
+40\,T^3I^2T_xT_{tx}J_{1x}-40\,I^2T_x^2T^2T_tJ_{1x}
+16\,I^2T^4X_{xx}XJ_{1x}+16\,T^4I^2X_x^2J_{1x}
\]
\[\fl\hspace{25pt}
+80\,I^2T_x^2UT^2J_{1x}-40\,I^2T_x^2T^2X_xJ_{1x}
-56\,T^2I^{2}T_{xxx}T_xJ_{1x}-32\,T^4J_{1x}I^6J_3
\]
\[\fl\hspace{25pt}
+8\,I^3J_{1x}T^{5}X_tJ_1+216\,I^2T_xT^2T_{xx}J_{1x}
+8\,T^3I^2T_{xxxx}J_{1x}+8\,I^3J_{1x}T^4XX_xJ_1
\]
\[\fl\hspace{25pt}
-36\,T^2I^2T_{xx}^2J_{1x}+16\,J_{1x}T^6II_{tt}
+16\,J_{1x}T^3I^2T_{xx}T_t-8\,J_{1x}T^4T_{tx}I^3J_1
\]
\[\fl\hspace{25pt}
-4\,I^3J_{1x}T_xT^3X^2J_1+15\,I^3J_{1x}T_x^3TJ_1
+4\,I^3J_{1x}T^3T_{xxx}J_1-18\,I^3J_{1x}T^2T_xT_{xx}J_1
\]
\[\fl\hspace{25pt}
-16\,I^3J_{1x}T_xUT^3J_1+16\,T^3I^2X_xT_{xx}J_{1x}
-16\,T^4I^2T_{txx}J_{1x}+8\,I^3J_{1x}T_xT^3X_xJ_1
\]
\[\fl\hspace{25pt}
-32\,I^2UT^3T_{xx}J_{1x}+16\,I^3J_{1x}T^4U_xJ_1+32\,T^4I^2U_{xx}J_{1x}
+8\,J_{1x}T_xT^3T_tI^3J_1
\]
\[\fl\hspace{25pt}
+16\,J_{1x}T^{5}I^2X_{tx})\,J_{1x}^{-1}T^{-4}I^{-6}J_1^{-1}.
\]
\noindent
They are invariants of the symmetry
pseudo-group for equation (\ref{para}) from $\Pc_2$.
The invariant differential ope\-ra\-tors are
\begin{equation}
\Dc_1 = {{\partial}\over{\partial \xi^1}} =
T\,I^{-1}\,D_t-J_{1t}\,I^{-2}\,J_{1x}^{-1}\,D_x,
\qquad
\Dc_2 = {{\partial}\over{\partial \xi^2}} = I^{-1}\,D_x,
\label{inv_diff_B}
\end{equation}
\noindent
where $D_t$ and $D_x$ are the operators of total differentiation w.r.t.
$t$ and $x$. These $\Dc_1$ and $\Dc_2$ are found without any integration.
Indeed, they satisfy $dF = \Dc_1(F)\,\xi^1 + \Dc_2(F)\,\xi^2$ for an arbitrary
function $F=F(t,x)$. Since $\xi^1 = I^2\,T^{-1}\,dt$
and $\xi^2 = I\,J_{1t}\,J_{1x}^{-1}\,dt+I\,dx$, we have (\ref{inv_diff_B}).

To construct all the other invariants of the pseudo-group, we apply $\Dc_1$
and $\Dc_2$ to $J_i$ in an arbitrary order:
$\Dc_1^{k_1}\Dc_2^{k_2}....\Dc_1^{k_{\beta-1}}\Dc_2^{k_{\beta}}\,J_i$.
The commutator identity
\[
\left[\Dc_1, \Dc_2\right] = J_1\,\Dc_1+J_2\,\Dc_2
\]
\noindent
allows us to permute the coframe derivatives, so we need only to deal with
the derived invariants
$J_{i,kl} = \Dc_1^{k}\Dc_2^{l}(J_i)$, $i\in\{1,...,4\}$, $k\ge 0$, and
$l\ge 0$. For $s\ge 0$ define the $s^{th}$ order {\it classifying manifold}
associated with the coframe $\bftheta = \{\theta_0, \theta_1, \theta_2, \xi^1,
\xi^2, \sigma_{11}, \sigma_{12}, \alpha_1, \alpha_2, \alpha_3\}$
and an open subset $U\subset\R^2$ as
\begin{equation}
\Cc^{(s)}(\bftheta,U) = \{(J_{i,kl}(t,x))\,\,\vert\,\,i\in\{1,...,4\},\,\,
k+l\le s, \,\,(t,x)\in U\}
\label{manifold_B}
\end{equation}
\noindent
Since all the functions $J_{i,kl}$ depend on two variables $t$ and $x$, it
follows that $\rho_s = \dim \Cc^{(s)}(\bftheta,U) \le 2$ for all $s\ge 0$.
Let $r=\min \{s \,\,\vert\,\, \rho_s = \rho_{s+1} = \rho_{s+2} = ...\}$
be the {\it order of the coframe} $\bftheta$. Since $J_{1x} \not = 0$, we have
$1\le \rho_0\le\rho_1\le\rho_2 \le ... \le 2$. In any case, $r+1 \le 2$.
Hence from Theorem 15.12 of \cite{Olver95} we see that
two linear parabolic equations (\ref{para}) from the subclass $\Pc_2$
are locally equivalent under a contact transformation if and only if
their second order classifying manifolds (\ref{manifold_B}) locally overlap.

\vskip 5 pt
\noindent
{\bf Remark} A Lie pseudo-group is called structurally intransitive,
\cite{LisleReid}, if it is not isomorphic to any transitive Lie pseudo-group.
In \cite{Cartan4}, Cartan proved that a Lie pseudo-group is structurally
intransitive whenever it has essential invariants. An invariant of a Lie
pseudo-group with the structure equations
\[
d\omega^i=A^i_{\beta k}\,\pi^{\beta}\w\omega^k+T^i_{jk}\,\omega^j\w\omega^k
\]
\noindent is called {\it essential}, if it is a first integral of the
{\it systatic system} $A^i_{\beta k}\,\omega^k$. From the structure equations
(\ref{SE_B}), it follows that the systatic system for the sym\-met\-ry
pseudo-group for an equation from $\Pc_2$ is generated by the forms $\xi^1$ and
$\xi^2$. First integrals of these forms are arbitrary functions of $t$ and $x$.
Therefore all the invariants $J_1$, ..., $J_4$, and all the derived invariants
are essential. Thus the symmetry pseudo-group of equation (\ref{para}) from the
subclass $\Pc_2$ is structurally intransitive.

\vskip 10 pt

Now we return to the case $J_{1x}=0$. Then the structure equations have the
form
\[\fl
d\theta_0=\alpha_1\w\theta_0
+\xi^1\w\theta_1
+\xi^2\w\theta_2,
\]
\[\fl
d\theta_1 = \alpha_1\w\theta_1+2\,\alpha_2\w\theta_2
-\case12\,J_1\,\alpha_2\w\theta_0
-\case12\,T^2\,I^{-4}\,J_{1t}\,(b^2_1-L_0)\,\theta_0\w\xi^1
\]
\[\fl\hspace{25pt}
+\case14\,T\,J_{1t}\,I^{-2}\,\theta_0\w\xi^2
+J_1\,\theta_1\w\xi^2
+\xi^1\w\sigma_{11}+\xi^2\w\sigma_{12},
\]
\[\fl
d\theta_2 =
\alpha_1\w\theta_2
+\alpha_2\w\theta_0
-\theta_1\w\xi^2
+\case12\,J_1\,\theta_2\w\xi^2
+\xi^1\w\sigma_{12},
\]
\[\fl
d\xi^1 = -J_1\,\xi^1\w\xi^2,
\]
\[\fl
d\xi^2 = - 2\,\alpha_2\w\xi^1,
\]
\[\fl
d\sigma_{11} = \alpha_1\w\sigma_{11}+4\,\alpha_2\w\sigma_{12}
-3\,J_1\,\alpha_2\w\theta_1
+\alpha_3\w\xi^1+\alpha_4\w\xi^2,
\]
\[\fl
d\sigma_{12} = \alpha_1\w\sigma_{11}+3\,\alpha_2\w\theta_1
-\case32\,J_1\,\alpha_2\w\theta_2+\alpha_4\w\xi^1-\theta_0\w\xi^1
\]
\[\fl\hspace{25pt}
-\case32\,T^2\,I^{-4}\,J_{1t}\,(b^2_1-L_0)\,\theta_2\w\xi^1
-\case34\,T\,I^{-2}\,J_{1t}\,(2\,\theta_1\w\xi^1-\theta_2\w\xi^2)
+2\,J_1\,\xi^1\w\sigma_{11}
\]
\[\fl\hspace{25pt}
+\xi^2\w\sigma_{11}
-\case32\,J_1\,\xi^2\w\sigma_{12},
\]
\noindent
where
\[\fl
L_0 =
-1/16\,(135\,{T_x}^{4}I^{2}
+16\,J_1T^{3}I^{3}T_xU
-16\,T^{5}I^{2}X_{tx}
+16\,T^{4}I^{2}X_{xxx}
-16\,J_1T^{4}I^{3}U_x
\]
\[\fl\hspace{25pt}
-8\,J_1T^{3}I^{3}T_xT_t
+40\,T^{2}I^{2}T_t{T_x}^{2}
-15\,J_1TI^{3}{T_x}^{3}
-4\,J_1T^{3}I^{3}T_xX^{2}
-20\,T^{2}I^{2}{T_x}^{2}X^{2}
\]
\[\fl\hspace{25pt}
-80\,T^{2}I^{2}{T_x}^{2}U
+8\,J_1T^{4}I^{3}X_{xx}
-16\,T^{4}I^{2}XX_{xx}+32\,T^{3}I^{2}UT_{xx}-216\,TI^{2}{T_x}^{2}T_{xx}
\]
\[\fl\hspace{25pt}
+18\,J_1T^{2}I^{3}T_xT_{xx}
+8\,T^{3}
I^{2}
X^{2}T_{xx}
-40\,T^{3}
I^{2}X_{xx}
T_x-16\,T^{4}
I^{2}{X_x}^{2}
-16\,T^{3}
I^{2}T_tT_{xx}
\]
\[\fl\hspace{25pt}
+56\,T^{2}
I^{2}
T_{xxx}T_x
+80\,T^{3}
I^{2}
U_xT_x
-4\,J_1T^{3}
I^{3}
T_{xxx}+32\,T^{6}
{I_t}^{2}+36\,T^{2}
I^{2}
{T_{xx}}^{2}
\]
\[\fl\hspace{25pt}
-8\,J_1T^{4}
I^{3}X
X_x
-8\,J_1T^{3}
I^{3}T_x
X_x+40\,T^{2}
I^{2}{T_x}^{2}
X_x-16\,T^{5}
IT_t
I_t+40\,T^{3}
I^{2}XX_xT_x
\]
\[\fl\hspace{25pt}
-16\,T^{3}
I^{2}X_x
T_{xx}-16\,T^{6}
I I_{tt}-8\,J_1
T^{5}
I^{3}X_t
+8\,T^{4}
I^{2}X_t
T_x-32\,T^{4}
I^{2}U_{xx}
\]
\[\fl\hspace{25pt}
+8\,J_1T^{4}
I^{3}T_{tx}
-40\,T^{3}
I^{2}T_{tx}
T_x
+16\,T^{4}
I^{2}
T_{txx}
-8\,T^{3}
I^{2}
T_{xxxx})\,
T^{-6}\,I^{-2}\,J_{1t}^{-1}.
\]

Consider the subclass $\Pc_3$ of all equations (\ref{para}) such that
$I\not=0$, $J_{1x}=0$, and $J_{1t}\not=0$. This subclass is not empty, since
the equation $u_{xx}=u_t + Q(t)\,x^{-2}\,u$ with $Q^{\prime}(t)\not = 0$
belongs to $\Pc_3$. For an equation from  $\Pc_3$, we normalize the
coefficient in the structure equation for $d\theta_1$ at $\theta_0\w\xi^1$
by setting $b^2_1=L_0$. Then we prolong the structure equations and obtain
\[\fl
d\theta_0=
\alpha_1\w\theta_0
+\xi^2\w\theta_2
+\xi^1\w\theta_1,
\]
\[\fl
d\theta_1=
\alpha_1\w\theta_1
-\case14\,J_1L_2\,\theta_0\,\w\xi^1
+\case14\,\left (\Dc_1(J_1)-J_1\,L_1\right )
\theta_0\w\xi^2
+J_1\,\theta_1\w\xi^2
+L_1\,\theta_2\w\xi^2
\]
\[\fl\hspace{25pt}
+\xi^2\w\sigma_{12}
+\xi^1\w\sigma_{11}
+L_2\,\theta_2\w\xi^1,
\]
\[\fl
d\theta_2=
\alpha_1\w\theta_2
+\case12\,L_2\,\theta_0\w\xi^1
+\case12\,L_1\,\theta_0\w\xi^2
-\theta_1\w\xi^2
+\case12\,J_1\,\theta_2\w\xi^2
+\xi^1\w\sigma_{12},
\]
\[\fl
d\xi^1=-J_1\,\xi^1\w\xi^2,
\]
\[\fl
d\xi^2=-L_1\,\xi^1\w\xi^2,
\]
\[\fl
d\sigma_{11}=
\alpha_1\w\sigma_{11}
+\alpha_2\w\xi^1
+\alpha_3\w\xi^2,
\]
\[\fl
d\sigma_{12}=
\alpha_1\w\sigma_{12}
+\alpha_3\w\xi^1
-\theta_0\w\xi^1
-\case32\left (\Dc_1(J_1)-J_1 L_1-L_2\right )\,\theta_1\w\xi^1
\]
\[\fl\hspace{25pt}
+\case32\,L_1\,\theta_1\w\xi^2
-\case34\,J_1L_2\,\theta_2\w\xi^1
+\case34\left (\Dc_1(J_1)-J_1 L_1\right )\,\theta_2\w\xi^2
+2\,J_1\,\xi^1\w\sigma_{11}
\]
\[\fl\hspace{25pt}
+2\,L_1\,\xi^1\w\sigma_{12}
+\xi^2\w\sigma_{11}
-\case32\,J_1\,\xi^2\w\sigma_{12},
\]
\[\fl
d\alpha_1=
\case14\,\left (2\,L_2-\Dc_1(J_1)+J_1L_1\right )
\xi^1\w\xi^2,
\]
\[\fl
d\alpha_2=
\pi_1\w\xi^1
+\pi_2\w\xi^2
+\alpha_1\w\alpha_2
-J_1\alpha_2\w\xi^2
+L_1\,\alpha_3\w\xi^2
\]
\[\fl\hspace{25pt}
-\case14\,\left (2\,L_2-\Dc_1(J_1)+J_1L_1\right )
\xi^2\w\sigma_{11},
\]
\[\fl
d\alpha_3=
\pi_2\w\xi^1
+\alpha_1\w\alpha_3
-\alpha_2\w\xi^2
+\case92\,J_1\,\alpha_3\w\xi^2
+\left (\case38\,{\Dc_1(J_1)}^{2}-\case34\,\Dc_1(J_1)J_1L_1
\right.
\]
\[\fl\hspace{25pt}
\left.
+\case38\,{J_1}^{2}{L_1}^{2}-\case52\,J_1\right )\,\theta_0\w\xi^2
-\case34\,(6\,\Dc_2(\Dc_1(J_1))-6\,\Dc_2(L_2)
-6\,J_1\Dc_2(L_1)+J_1\Dc_1(J_1)
\]
\[\fl\hspace{25pt}
-6\,J_1 L_2
-J_1^2 L_1
-2\,{L_1}^{2}
+6\,\Dc_1(L_1)
)\,
\theta_1\w\xi^2
+\left (\case92\,L_1\Dc_1(J_1)
-\case{15}{4}\,J_1{L_1}^{2}
-\case32\,{J_1}^{2}L_2
\right.
\]
\[\fl\hspace{25pt}
\left.
-1+\case34\,J_1\Dc_1(L_1)-\case34\,J_1\,\Dc_2(L_2)
-\case34\,\Dc_1^2(J_1)\right )\,
\theta_2\w\xi^2
+\left (\case92\,L_1+5\,{J_1}^{2}\right )\,\xi^2\w\sigma_{11}
\]
\[\fl\hspace{25pt}
-2\,(L_2-2\Dc_1(J_1)+J_1L_1-\Dc_2(L_1))\,\xi^2\w\sigma_{12},
\]
\noindent
where
\[\fl
L_1 = T\,I^{-3}\,(L_{0x}-I_t),
\]
\[\fl
L_2=-1/8\,(8\,I^{2}
T^{3}
T_{tx}
-8\,I^{2}
T_x
T^{2}
X_x
-15\,I^{2}
T_x^{3}
+4\,I^{2}
T_x
T^{2}
X^{2}
+16\,I^{2}
T_xU
T^{2}-8\,T^{4}
I^{2}X_t
\]
\[\fl\hspace{25pt}
-8\,I^{2}
T_x
T^{2}T_t
-8\,IT^{5}
L_{0t}
+18\,TI^{2}
T_x
T_{xx}
-8\,L_0T^{4}I
T_t
-4\,L_0^{2}T^{5}
IJ_1+8\,T^{4}L_0L_1
I^{3}
\]
\[\fl\hspace{25pt}
+16\,T^{5}
I_tL_0
+8\,I^{2}T^{3}
X_{xx}
-8\,I^{2}T^{3}X
X_x-4\,T^{2}
I^{2}T_{xxx}
-16\,I^{2}T^{3}
U_x)
\,I^{-5}
T^{-3},
\]
\noindent
and the invariant differential operators are defined by
\[
\Dc_1 = {{\partial}\over{\partial \xi^1}} =
T\,I^{-2}\,D_t-T\,I^{-3}\,L_0\,D_x,
\qquad
\Dc_2 = {{\partial}\over{\partial \xi^2}} = I^{-1}\,D_x.
\]
\noindent
The commutator relation for invariant differentiations is
\[
\left[\Dc_1, \Dc_2\right] = J_1\,\Dc_1+L_1\,\Dc_2.
\]
\noindent
The $s^{th}$ order classifying manifold associated with the involutive coframe
$\bftheta =\{ \theta_0, \theta_1, \theta_2, \xi^1, \xi^2, \sigma_{11},
\sigma_{12}, \alpha_1, \alpha_2, \alpha_3\}$ and an open subset $U\subset\R^2$
is
\begin{equation}\fl
\Cc^{(s)}(\bftheta,U) = \{\left(
\Dc_1^k\Dc_2^l(J_1(t,x)),\,
\Dc_1^k\Dc_2^l(L_i(t,x))\right)\,
\,\,
\vert\,\,i\in\{1,2\},\,\, k+l\le s,
\,\,(t,x)\in U\}
\label{manifold_C}
\end{equation}
\noindent
Thus
two linear parabolic equations (\ref{para}) from the subclass $\Pc_3$
are locally equivalent under a contact transformation if and only if
their second order classifying manifolds (\ref{manifold_C}) locally overlap.

Since all the invariants of the symmetry pseudo-group for an equation from
$\Pc_3$ are first integrals of the systatic system $\xi^1$, $\xi^2$, this
pseudo-group is structurally intransitive.

\vskip 5 pt

Now we return to the case $J_{1x} = J_{1t} = 0$. We denote
$J_1=N=const$, then the structure equations have the
form
\[\fl
d \theta_0=
\alpha_1\w\theta_0
+\xi^1\w\theta_1
+\xi^2\w\theta_2,
\]
\[\fl
d \theta_1=
\alpha_1\w\theta_1
-\case12\,N\,\alpha_2\w\theta_0
+2\,\alpha_2\w\theta_2
+M_1\,\theta_0\w\xi^1
+N\,\theta_1\w\xi^2
+\xi^1\w\sigma_{11}
+\xi^2\w\sigma_{12},
\]
\[\fl
d \theta_2=
\alpha_1\w\theta_2
+\alpha_2\w\theta_0
-\theta_1\w\xi^2
+\case12\,N\,\theta_2\w\xi^2
+\xi^1\w\sigma_{12},
\]
\[\fl
d \xi^1=-N\,\xi^1\w\xi^2,
\]
\[\fl
d \xi^2=-2\,\alpha_2\w\xi^1,
\]
\[\fl
d\sigma_{11}=
\alpha_1\w\sigma_{11}
-3\,N\,\alpha_2\w\theta_1
+4\,\alpha_2\w\sigma_{12}
+\alpha_3\w\xi^1
+\alpha_4\w\xi^2,
\]
\[\fl
d\sigma_{12}=
\alpha_1\w\sigma_{12}
+3\,\alpha_2\w\theta_1
-\case32\,N\,\alpha_2\w\theta_2
+\alpha_4\w\xi^1
-\theta_0\w\xi^1
+3\,M_1\,\theta_2\w\xi^1
\]
\[\fl\hspace{25pt}
+2\,N\,\xi^1\w\sigma_{11}
+\xi^2\w\sigma_{11}
-\case32\,N\,\xi^2\w\sigma_{12},
\]
\noindent
where
\[\fl
M_1=1/32\,(-40\,
I^{2}
T^{3}X
X_{{x}}T_{{x}}
+32\,I^{2}
T^{4}U_{xx}
-32\,T^{6}{I_t}^{2}
+16\,N\,I^{3}
T^{4}U_{{x}}
+8\,I^{2}
T^{3}
T_{xxxx}
\]
\[\fl\hspace{25pt}
+16\,I^{2}
T^{4}
XX_{xx}
-8\,N\,I^{3}
T^{4}X_{xx}
+8\,N\,I^{3}
T^{4}
XX_{{x}}
+8\,N\,I^{3}
T^{5}X_{{t}}
+16\,I^{2}
T^{5}X_{tx}
\]
\[\fl\hspace{25pt}
-8\,N\,I^{3}
T^{4}T_{tx}
+16\,T^{5}I
T_{{t}}I_t
+80\,I^{2}
T^{2}{T_{{x}}}^{2}
U+20\,I^{2}
T^{2}{T_{{x}}}^{2}
X^{2}-16\,N\,
I^{3}
T^{3}T_{{x}}
U
\]
\[\fl\hspace{25pt}
+15\,N\,
I^{3}T
{T_{{x}}}^{3}
-56\,I^{2}
T^{2}T_{xxx}
T_{{x}}+216\,I^{2}
T{T_{{x}}}^{2}
T_{xx}+40\,I^{2}
T^{3}X_{xx}
T_{{x}}
\]
\[\fl\hspace{25pt}
-18\,N\,I^{3}
T^{2}T_{{x}}
T_{xx}
+16\,I^{2}
T^{4}{X_{{x}}}^{2}
+40\,I^{2}
T^{3}T_{tx}
T_{{x}}-135\,I^{2}
{T_{{x}}}^{4}+8\,N\,
I^{3}
T^{3}T_{{x}}
X_{{x}}
\]
\[\fl\hspace{25pt}
-8\,I^{2}
T^{4}X_{{t}}
T_{{x}}-40\,I^{2}
T^{2}{T_{{x}}}^{2}
X_{{x}}-80\,I^{2}
T^{3}U_{{x}}
T_{{x}}-8\,I^{2}
T^{3}X^{2}
T_{xx}-32\,I^{2}
T^{3}U
T_{xx}
\]
\[\fl\hspace{25pt}
+16\,I^{2}
T^{3}X_{{x}}
T_{xx}+4\,N\,
I^{3}
T^{3}T_{xxx}
+16\,I^{2}
T^{3}
T_{{t}}T_{xx}
+8\,N\,I^{3}
T^{3}T_{{x}}
T_{{t}}-40\,I^{2}
T^{2}
T_{{t}}{T_{{x}}}^{2}
\]
\[\fl\hspace{25pt}
-4\,N\,I^{3}
T^{3}T_{{x}}
X^{2}
-36\,I^{2}
T^{2}{T_{xx}}^{2}
-16\,I^{2}
T^{4}
X_{xxx}-16\,I^{2}
T^{4}T_{{txx}}
+16\,IT^{6}
I_{tt})
\,I^{-6}\,T^{-4}.
\]
\noindent
All the essential torsion coefficients now are independent of the group
parameters, but
\[
d M_1 = \left((\case32\,N\,M_1+1)\,b^2_1+M_{1t}\right)\,T\,I^{-2}\,\xi^1
-\left(\case32\,M_1\,N-1\right)\,\xi^2.
\]

By $\Pc_4$ we denote the subclass  of all equations (\ref{para}) such that
$I\not = 0$, $J_1 = N = const$, and $3\,N\,M_1\not = -2$. This subclass
contains, e.g., the equation $u_{xx}=u_t+(\kappa\,x^{-2}+\nu\,x)\,u$ with
$\kappa \not=0$, $\nu \not=0$. For an equation from $\Pc_4$, we set
$b^2_1 = -2\,M_{1t}\,(3\,N\,M_1+2)^{-1}$. After this
normalization, we prolong the structure equations and obtain
\[\fl
d\theta_0=
\alpha_1\w\theta_0
+\xi^1\w\theta_1
+\xi^2\w\theta_2,
\]
\[\fl
d\theta_1=
\alpha_1\w\theta_1
+\left (M_1-N\,M_3\right )\,\theta_0\w\xi^1
-\case14\,N\,M_2\,\theta_0\w\xi^2
+N\,\theta_1\w\xi^2
+4\,M_3\,\theta_2\w\xi^1
\]
\[\fl\hspace{25pt}
+M_2\,\theta_2\w\xi^2
+\xi^1\w\sigma_{11}
+\xi^2\w\sigma_{12},
\]
\[\fl
d\theta_2=
\alpha_1\w\theta_2
+2\,M_3\,\theta_0\w\xi^1
+\case12\,M_2\,\theta_0\w\xi^2
-\theta_1\w\xi^2
+\case12\,N\,\theta_2\w\xi^2
+\xi^1\w\sigma_{12},
\]
\[\fl
d\xi^1=
-N\,\xi^1\w\xi^2,
\]
\[\fl
d\xi^2=
-M_2\,\xi^1\w\xi^2,
\]
\[\fl
d\sigma_{11}=
\alpha_1\w\sigma_{11}
+\alpha_2\w\xi^1
+\alpha_3\w\xi^2,
\]
\[\fl
d\sigma_{12}=
\alpha_1\w\sigma_{12}
+\alpha_3\w\xi^1
-\theta_0\w\xi^1
+\left (6\,M_3+\case32\,N\,M_2\right )\,\theta_1\w\xi^1
+\case32\,M_2\,\theta_1\w\xi^2
\]
\[\fl\hspace{25pt}
+3\,\left (M_1-N\,M_3\right )\,\theta_2\w\xi^1
-\case34\,N\,M_2\,\theta_2\w\xi^2
+2\,N\,\xi^1\w\sigma_{11}
+2\,M_2\,\xi^1\w\sigma_{12}
\]
\[\fl\hspace{25pt}
+\xi^2\w\sigma_{11}
-\case32\,N\,\xi^2\w\sigma_{12},
\]
\[\fl
d\alpha_1=
\left (2\,M_3+\case14\,N\,M_2\right )\,
\xi^1\w\xi^2,
\]
\[\fl
d\alpha_2=
\pi_1\w\xi^1
+\pi_2\w\xi^2
+\alpha_1\w\alpha_2
+N\,\alpha_2\w\xi^2
+M_2\,\alpha_3\w\xi^2
- (2\,M_3+\case14\,N\,M_2)\,\xi^2\w\sigma_{11},
\]
\[\fl
d\alpha_3=
\pi_2\w\xi^1
+\alpha_1\w\alpha_3
-\alpha_2\w\xi^2
+\case92\,N\,\alpha_3\w\xi^2
+\left (\case38\,{N}^{2}{M_2}^{2}-\case52\,N\right )\,\theta_0\w\xi^2
\]
\[\fl\hspace{25pt}
+ (3\,M_1-\case32\,\Dc_1(M_2)+6\,N\,M_3
+\case94\,{N}^{2}M_2+\case92\,{M_2}^{2}
+\case32\,N\,\Dc_2(M_2)
\]
\[\fl\hspace{25pt}
+6\,\Dc_2(M_3) )\,\theta_1\w\xi^2
+\left (6\,N\,M_1-\case{15}{4}\,N\,{M_2}^{2}
-6\,{N}^{2}M_3-1+3\,\Dc_2(M_1)
\right.
\]
\[\fl\hspace{25pt}
\left.
+\case34\,N\,\Dc_1(M_2)
+3\,N\,\Dc_2(M_3)\right )\,\theta_2\w\xi^2
+\left (\case92\,M_2+5\,{N}^{2}\right )\,\xi^2\w\sigma_{11}
\]
\[\fl\hspace{25pt}
-\left (8\,M_3+2\,N\,M_2-2\,\Dc_2(M_2)\right )\,\xi^2\w\sigma_{12},
\]
\noindent
where
\[\fl
M_2= T\,\left (M_{0x}-I_t\right )\,I^{-3},
\qquad
M_0=-2\,M_{1t}\,(3\,N\,M_1+2)^{-2},
\]
\[\fl
M_3=-1/32\,(-8\,IT^5M_{0t}
-8\,T^4M_0M_2I^3
+16\,T^5I_tM_0
-4\,M_0^2T^5IN
-16\,I^2T^3U_x
\]
\[\fl\hspace{25pt}
+8\,I^2
T^3X_{xx}
+8\,I^2T^3
T_{tx}-4\,T^2
I^2T_{xxx}
+16\,I^2T_{x}
UT^2-8\,M_0T^4I
T_{t}
\]
\[\fl\hspace{25pt}
-8\,I^2
T^3XX_{x}
+18\,TI^2
T_{x}
T_{xx}+4\,I^2
T_{x}
T^2X^2-8\,I^2
T_{x}T^2
T_{t}-8\,I^2
T_{x}T^2
X_{x}
\]
\[\fl\hspace{25pt}
-15\,I^2
T_{x}^3-8\,T^4
I^2X_t)\,
I^{-5}\,T^{-3}.
\]
The invariant differential operators
\[
\Dc_1 = T\,I^{-2}\,D_t-T\,M_0\,I^{-3}\,D_x,
\qquad
\Dc_2 = I^{-1}\,D_x,
\]
\noindent
satisfy the commutator relation
\[
\left[\Dc_1, \Dc_2\right] = N\,\Dc_1+M_2\,\Dc_2.
\]
\noindent
The $s^{th}$ order classifying manifold associated with the involutive coframe
$\bftheta =\{ \theta_0, \theta_1, \theta_2, \xi^1, \xi^2, \sigma_{11},
\sigma_{12}, \alpha_1, \alpha_2, \alpha_3\}$ and an open subset $U\subset\R^2$
is
\begin{equation}\fl
\Cc^{(s)}(\bftheta,U) = \{
(\Dc_1^k\Dc_2^l(M_i(t,x)))
\,\,
\vert\,\,i\in\{1,2,3\},\,\, k+l\le s,
\,\,(t,x)\in U\}.
\label{manifold_D}
\end{equation}
\noindent
So two linear parabolic equations (\ref{para}) from the subclass $\Pc_4$
are locally equivalent under a contact transformation if and only if
their second order classifying manifolds (\ref{manifold_D}) locally overlap.

The systatic system for the symmetry pseudo-group of equation (\ref{para})
from the subclass $\Pc_4$ is generated by  $\xi^1$ and $\xi^2$ again, and,
as all the differential invariants are essential, this pseudo-group is
structurally intransitive.

\vskip 5 pt

Finally, consider the subclass $\Pc_5$ of all equations (\ref{para}) such that
$I\not = 0$, $J_1 =N = const$, and $M_1 = -2/(3\,N)$. For an equation
from $\Pc_5$, after a prolongation, the structure equations have the form
\[\fl
d\theta_0=
\alpha_1\w\theta_0
+\xi^1\w\theta_1
+\xi^2\w\theta_2,
\]
\[\fl
d\theta_1=
\alpha_1\w\theta_1
-\case{N}2\,\alpha_2\w\theta_0
+2\,\alpha_2\w\theta_2
-\case2{3N}\,\theta_0\w\xi^1
+N\,\theta_1\w\xi^2
+\xi^1\w\sigma_{11}
+\xi^2\w\sigma_{12},
\]
\[\fl
d\theta_2=
\alpha_1\w\theta_2
+\alpha_2\w\theta_0
-\theta_1\w\xi^2
+\case{N}2\,\theta_2\w\xi^2
+\xi^1\w\sigma_{12},
\]
\[\fl
d\xi^1=-N\,\xi^1\w\xi^2,
\]
\[\fl
d\xi^2=-2\,\alpha_2\w\xi^1,
\]
\[\fl
d\sigma_{11}=
\alpha_1\w\sigma_{11}
-3\,N\,\alpha_2\w\theta_1
+4\,\alpha_2\w\sigma_{12}
+\alpha_3\w\xi^1
+\alpha_4\w\xi^2,
\]
\[\fl
d\sigma_{12}=
\alpha_1\w\sigma_{12}
-\case{3N}{2}\,\alpha_2\w\theta_2
+3\,\alpha_2\w\theta_1
+\alpha_4\w\xi^1
-\theta_0\w\xi^1
-\case2N\,\theta_2\w\xi^1
+2\,N\,\xi^1\w\sigma_{11}
\]
\[\fl\hspace{25pt}
+\xi^2\w\sigma_{11}
-\case{3N}{2}\,\xi^2\w\sigma_{12},
\]
\[\fl
d\alpha_1=
\case{N}2\,\alpha_2\w\xi^1
-\alpha_2\w\xi^2,
\]
\[\fl
d\alpha_2=
N\,\alpha_2\w\xi^2
-\case{2}{3N}\,\xi^1\w\xi^2,
\]
\[\fl
d\alpha_3=
\pi_1\w\xi^1
+\pi_2\w\xi^2
+\alpha_1\w\alpha_3
+6\,\alpha_2\w\alpha_4
-2\,\alpha_2\w\theta_0
-\case{8}{N}\,\alpha_2\w\theta_2
-\case{9N}2\,\alpha_2\w\sigma_{11}
\]
\[\fl\hspace{25pt}
+N\,\alpha_3\w\xi^2
+2\,\theta_1\w\xi^2
+\case{8}{3N}\,\xi^2\w\sigma_{12},
\]
\[\fl
d\alpha_4=
\pi_2\w\xi^1
+\alpha_1\w\alpha_4
-6\,{N}^{2}\alpha_2\w\theta_1
-5\,\alpha_2\w\sigma_{11}
+13\,N\,\alpha_2\w\sigma_{12}
-\alpha_3\w\xi^2
\]
\[\fl\hspace{25pt}
+\case{9N}{2}\,\alpha_4\w\xi^2
-\case{5N}{2}\,\theta_0\w\xi^2
-\case{4}{N}\,\theta_1\w\xi^2
-4\,\theta_2\w\xi^2
+5\,{N}^{2}\xi^2\w\sigma_{11}.
\]
\noindent
From these structure equations, it follows that the classifying manifold is a
point, and that two equations from the subclass $\Pc_5$ are equivalent
under a contact transformation iff they have the same value of the constant
$N$. Repeating the calculations for the equation
\begin{equation}
u_{xx}=u_{t}+ \tilde{N}\,x^{-2}\,u,
\label{Nu}
\end{equation}
\noindent
we see that its symmetry pseudo-group has the same structure equations
whenever $\tilde{N} = - 4/(3\,N^5)$. Thus
the linear parabolic equation (\ref{para}) is equivalent to an equation
of the form (\ref{Nu}) under a contact transformation if and only if
it belongs to the subclass $\Pc_5$.

\vskip 5 pt
The results of the calculations are summarized in the following statement:

\vskip 5 pt
\noindent{\bf Theorem 2}
{\it The class of linear parabolic equations (\ref{para}) is divided into
the five subclasses} $\Pc_1$, $\Pc_2$, ..., $\Pc_5$ {\it invariant under an
action of the pseudo-group of contact transformations:}

$\Pc_1$ {\it consists of all equations (\ref{para}) such that} $I = 0$;

$\Pc_2$ {\it consists of all equations (\ref{para}) such that} $I \not= 0$
{\it and} $J_{1x}\not = 0$;

$\Pc_3$ {\it consists of all equations (\ref{para}) such that} $I \not= 0$,
$J_{1x} = 0$, {\it and}
$J_{1t}\not = 0$;

$\Pc_4$ {\it consists of all equations (\ref{para}) such that} $I \not= 0$,
$J_1 =N= const$, {\it and}
$3\,N\,M_1 \not = -2$;

$\Pc_5$ {\it consists of all equations (\ref{para}) such that} $I \not= 0$,
$J_1 =N= const$, {\it and}
$3\,N\,M_1 = -2$.

{\it Every equation from the subclass} $\Pc_1$ {\it is equivalent to
the linear heat equation (\ref{heat}).}

{\it Two equations from one of the subclasses} $\Pc_2$, $\Pc_3$,
{\it or} $\Pc_4$ {\it are locally equivalent to each other if and
only if the classifying manifolds (\ref{manifold_B}),
(\ref{manifold_C}), or (\ref{manifold_D}) for these equations
locally overlap.}

{\it Every equation from the subclass} $\Pc_5$ {\it is locally equivalent
to the equation (\ref{Nu}) whenever}
$\tilde{N} = - 4/(3\,N^5)$.

\section*{Conclusion}
In this paper, the moving coframe method of \cite{FO} is applied
to the local equivalence problem for the class of linear
second-order parabolic equations in two independent variables
under an action of the pseudo-group of contact transformations.
The class is divided into the five invariant subclasses. We have
found the structure equations and the complete sets of
differential equations for all the subclasses. The solution of the
equivalence problem is given in terms of the differential
invariants. It is shown that the moving coframe method is
applicable to structurally intransitive symmetry pseudo-groups.
The moving coframe method allows us to find invariant 1-forms,
structure equations, differential invariants, and operators of
invariant differentiation for symmetry pseudo-groups of
differential equations without analyzing over-determined systems
of partial differential equation or using procedures of
integration.


\section*{References}

\begin{thebibliography}{99}
\bibitem{Cartan1} Cartan \'E 1953 {\it  Sur la structure des groupes infinis
        de transformations //
        {\OE}uvres Compl{\`e}tes}, Part II,  {\bf  2}
        (Paris: Gauthier - Villars)
        571--714
\bibitem{Cartan2} Cartan \'E 1953 {\it
        Les sous-groupes des groupes continus de transformations //
        {\OE}uvres Compl{\`e}tes}, Part II,  {\bf  2}
        (Paris: Gauthier - Villars)
        719--856
\bibitem{Cartan3} Cartan \'E 1953 {\it
        Les groupes de transformations continus, infinis, simples //
        {\OE}uvres Compl{\`e}tes}, Part II,  {\bf  2}
        (Paris: Gauthier - Villars)
        857--925
\bibitem{Cartan4} Cartan \'E 1953 {\it
        La structure des groupes infinis. //
        {\OE}uvres Compl{\`e}tes}, Part II,  {\bf  2}
        (Paris: Gauthier - Villars)
        1335--84
\bibitem{Cartan5} Cartan \'E 1953 {\it
        Les probl\`emes d'\'equivalence. //
        {\OE}uvres Compl{\`e}tes}, Part II,  {\bf  2}
        (Paris: Gauthier - Villars)
        1311--1334
\bibitem{FO} Fels M, Olver P J 1998 {\it Moving coframes. I. A practical
        algorithm // Acta. Appl. Math} {\bf 51} 161--213
\bibitem{FO2} Fels M, Olver P J 1999 {\it Moving coframes. II. Regularization
        and theoretical foundations // Acta Appl. Math}  {\bf 55} 127--208
\bibitem{Gardner} Gardner R B 1989 {\it The Method of Equivalence and Its
        Applications} (Philadelphia: SIAM)
\bibitem{Ibragimov2002} Ibragimov N H 2002 {\it Laplace Type Invariants
        for Parabolic Equations // Nonlinear Dynamics} {\bf 28} 125 --133
\bibitem{JM} Johnpillai I K, Mahomed F M 2001
        {\it Singular invariant equation for the (1+1) Fokker - Planck
        equation // J Phys A Math Gen} {\bf 34} 11033--11051
\bibitem{Lie} Lie S 1922 - 1937 {\it Gesammelte Abhandlungen} {\bf 1 - 6}
         (Leipzig: B G Teubner)
\bibitem{LisleReid}  Lisle I G, Reid G J  1998 {\it Geometry and structure of
        Lie pseudogroups from infinitesimal defining equations. //
        J. Symb. Comp.} {\bf 26} 355--79
\bibitem{Morozov} Morozov O I  2002
        {\it Moving Coframes and Symmetries of Differential Equations. //
        J Phys A Math Gen} {\bf 35} 2965 -- 2977
\bibitem{Ovsiannikov} Ovsiannikov L V 1982 {\it Group Analysis of Differential
        Equations} (New York: Academic Press)
\bibitem{Olver95} Olver P J 1995 {\it Equivalence, Invariants, and Symmetry}
        (Cambridge: Cambridge University Press)
\bibitem{Stohny} Stohny V 1997
        {\it Symmetry Properties and Exact Solutions of the Fokker-Planck
        Equation.//
        Nonlinear Mathematical Physics} {\bf 4} N 1--2 pp 132 -- 136
\end{thebibliography}
\end{document}